\documentclass{aa}
\usepackage{graphicx}
\usepackage{txfonts}
\usepackage[]{natbib}
\titlerunning{Cold dust and molecular gas content in Virgo spirals}

\begin{document}
   \title{The Herschel Virgo Cluster Survey: X.The relationship between cold dust and molecular gas content in 
         Virgo spirals\thanks{{\it Herschel} is an ESA space observatory with science instruments provided by 
          European-led  Principal Investigator consortia and with important participation from NASA.}}

   \author{Edvige Corbelli
          \inst{1}
         \and
          Simone Bianchi
          \inst{1}
         \and
          Luca Cortese
          \inst{2}
  	 \and
          Carlo Giovanardi
          \inst{1}
         \and
          Laura Magrini
          \inst{1}
	 \and
	  Ciro Pappalardo
	  \inst{1}
 	 \and      
          Alessandro Boselli
          \inst{3}
	 \and      
          George J. Bendo
          \inst{4}
         \and
          Jonathan Davies
          \inst{5}
          \and
          Marco Grossi
          \inst{6}
	 \and
	  Suzanne C. Madden
	 \inst{7}
         \and
          Matthew W. L. Smith
          \inst{5}
 	 \and      
          Catherine Vlahakis
          \inst{8}
	 \and
	  Robbie Auld
	  \inst{5}
	 \and
          Maarten Baes
          \inst{9}
         \and
          Ilse De Looze
          \inst{9}
         \and
          Jacopo Fritz
          \inst{9}
         \and
          Michael Pohlen
          \inst{5}
	 \and
          Joris Verstappen
          \inst{9}
 	}

%   \offprints{}

   \institute{
Osservatorio Astrofisico di Arcetri - INAF, Largo E. Fermi 5, 50125
Firenze, Italy
\email{[edvige, sbianchi, giova, laura, cpappala]@arcetri.astro.it}
\and
European Southern Observatory, Karl-Schwarzschild-Strasse 2, D-85748 Garching bei Munchen, Germany
\and
Laboratoire d'Astrophysique de Marseille, UMR 6110 CNRS, 38 rue F. Joliot-Curie, F-13388 Marseille, France
\and
Jodrell Bank Centre for Astrophysics, Alan Turing Building, School of Physics and Astronomy, University of Manchester, Manchester M13 9PL
\and
Department of Physics and Astronomy, Cardiff University, The Parade, Cardiff, CF24 3AA, UK
\and
CAAUL, Observatorio Astronomico de Lisboa, Universidade de Lisboa, Tapada de Ajuda, 1349-018, Lisboa, Portugal
\and
Service d'Astrophysique, CEA/Saclay, l'Orme des Merisiers, 91191 Gif-sur-Yvette, France
\and
Joint ALMA Office, Alonso de Cordova 3107, Vitacura,  and
Departamento de Astronomia, Universidad de Chile, Casilla 36-D, Santiago, Chile
\and
Sterrenkundig Observatorium, Universiteit Gent, Krijgslaan 281 S9, B-9000 Gent, Belgium
          }
   \date{Received; accepted}

  \abstract
   {}
    {We examine whether dust mass can trace the total or molecular gas mass in late-type   
    Virgo cluster galaxies, and how the environment affects the dust-to-gas ratio and the molecular fraction.}  
   {Using the far-infrared emission, as observed by the  ${Herschel}$
   Virgo Cluster Survey (HeViCS), and the integrated HI and CO brightness, we infer the dust and total gas mass 
   for a magnitude limited sample of 35 metal rich spiral galaxies. Environmental disturbances on each 
   galaxy are considered by means of the HI deficiency parameter.}  
   {The CO flux correlates tightly and linearly with far-infrared fluxes observed by Herschel,  
   especially with the emission at 160, 250 and 350$\mu$m. Molecules in these 
   galaxies are more closely related to cold dust rather than to 
   dust heated by star formation or to optical/NIR brightness. We show that dust mass establishes a stronger
   correlation with the total gas mass than with the atomic or molecular component alone. 
   The correlation is non-linear since lower mass galaxies have a 
   lower dust-to-gas ratio. The dust-to-gas ratio increases as the HI deficiency increases, but in
   highly HI deficient galaxies it stays constant. Dust is in fact less affected than atomic gas 
   by weak cluster interactions, which remove most of the HI gas from outer and high latitudes regions. 
   Highly disturbed galaxies, in a dense cluster environment, can instead loose a considerable fraction of gas 
   and dust from the inner regions of the disk keeping constant 
   the dust-to-gas ratio. There is evidence that the molecular phase is also quenched. This quencing becomes evident 
   by considering the molecular gas mass per unit stellar mass. Its amplitude, if confirmed by future studies, 
   highlights that molecules are missing in Virgo HI deficient spirals, but to a somewhat lesser extent than dust.
   }
   {}

   \maketitle
%
%________________________________________________________________

%\tableofcontents

\section{Introduction}

Dust and gas in galaxies are  essential ingredients for star
formation. Stars are born because of the cooling and fragmentation
of molecular gas, which in today's galaxies forms from atomic gas primarily 
by catalytic reactions on the surface of 
dust grains \citep{1963ApJ...138..393G}. Dust and atomic gas are, however, only part 
of the picture because the 
fractional abundance of molecules depends on a delicate balance between 
formation and dissociation processes \citep{1971ApJ...163..165H}
and on the self-shielding ability of molecular species from ultraviolet 
radiation \citep{1996ApJ...468..269D}. The presence of prominent arms and high
interstellar pressure in the disk favour high molecular abundances \citep{1993ApJ...411..170E}. 
In order to study the relation between dust and molecules it is important to have 
a good inventory of their masses and distribution in a variety of galaxy types and environments. 
Intense interstellar radiation fields and galaxy-galaxy interactions 
or gas stripping as the galaxy moves through inter-cluster gas can  
quench the growth of molecular clouds through galaxy starvation or strangulation
\citep{1980ApJ...237..692L,2002ApJ...577..651B,2006PASP..118..517B,2008A&A...490..571F,2009ApJ...697.1811F}. 
But there is not a unanimous consent on how environmental conditions affect the molecular component:
this is more bounded to the galaxy potential well, being more confined inside the disk, and
less disturbed by the environment
\citep{1986ApJ...310..660S,1988ApJ...334..613S,1989ApJ...344..171K}. 

Measuring the total molecular content of a galaxy is a rather
difficult task because molecular hydrogen (H$_2$), the most abundant molecular  
species, is hard to excite, lacks a permanent electric dipole moment and
must radiate through a slow quadrupole transition.
The ability of molecular hydrogen to form stars, the so called star formation efficiency, can
also vary and hence the star formation rate cannot be used as an
indicator of molecular mass \citep{1999ApJ...514L..87Y,2007A&A...468...61D,2011ApJ...729..133M}.
Molecular surveys rely on the brightness of CO rotational lines
and on CO-to-H$_2$ conversion factors, X$_{CO}$, to derive the molecular hydrogen mass from  
the CO luminosity. 
In general this depends on the CO-to-H$_2$ abundance ratio, which
is a function of metallicity and on the self-shielding ability of the two molecules,
and on the excitation state of the CO molecules.    
The CO molecule forms almost exclusively in the gas  phase and has a lower capability for 
self-shielding then H$_2$ due to the low abundance of 
carbon relative to hydrogen \citep[e.g.][]{2011MNRAS.412..337G}. 
However, there are still discrepant results on how X$_{CO}$ 
varies, for example as a function of the gas metallicity 
\citep{1995ApJ...448L..97W,1997A&A...328..471I,2008ApJ...686..948B,2011ApJ...737...12L,
2011MNRAS.415.3253S}. 
Infrared emission has been used sometime to find possible variations of the CO-to-H$_2$ conversion
factor \citep{1987AJ.....94...54K,1997A&A...317...65I,1997A&A...328..471I,2007ApJ...658.1027L,
2010A&A...512A..68G,2011ApJ...737...12L,2011A&A...535A..13M} and estimate the molecular content. 

The total dust mass of a galaxy can only be determined
if dust emission is traced throughout using multi-wavelength observations
from the infrared (hereafter IR) to the sub-millimeter. Hot dust,
in the proximity of star-forming regions, is similar to the
tip of an iceberg: prominent but not representative of the total dust mass.
The {\it Herschel Space Observatory} \citep{2010A&A...518L...1P,2010A&A...518L...2P, 
2010A&A...518L...3G}, with its two photometric instruments
at far infrared wavelengths (PACS, the Photodetector Array Camera Spectrometer with 70,
100, and 160~$\mu$m\ bands and SPIRE, the Spectral and Photometric Imaging Receiver with
250,350 and 500~$\mu$m\ bands) can observe both sides of the peak ($100-200\mu$m) in the  
spectral energy distribution of dust in galaxies.
The {\it Herschel} Virgo Cluster Survey \citep[][hereafter HeViCS ]{2010A&A...518L..48D}, an 
{\it Herschel} Open Time Key Project, has
mapped an area of the nearby Virgo cluster of galaxies,
with PACS and SPIRE to investigate the dust content of galaxies of different morphological 
types and whether intra-cluster dust is present. Recent studies 
based on Herschel data \citep{2010A&A...518L..49C,2012A&A...540A..52C} have already pointed out
that in a dense cluster environment dust can be removed from galaxy disks. 

For a sample of bright spiral galaxies in the Virgo cluster, which have been targets of CO observations,
we shall examine  the correlation between the dust emission and the CO brightness, 
integrated over the disk, and  between dust and gas masses. An important question 
to address is, for example, whether HI deficient galaxies are also deficient in their global dust and 
molecular content and how deficiency affects galaxy evolution \citep{2006PASP..118..517B}.
The large number of Virgo spiral galaxies, forming 
stars actively, will let us investigate whether dust mass establishes a better correlation with the 
molecular gas mass or with the sum of the atomic and molecular gas mass and to pin down possible 
environmental effects on the dust-to-gas ratio.  
A large sample of field galaxies would be an ideal reference sample to have,
in order to distinguish environmental effects from properties related to the galaxy quiescent evolution. 
However, the presence of newly acquired cluster members 
in a young cluster, such as the Virgo Cluster, offers the possibility of using the HI deficiency 
as indicator of environmental disturbances. By defining the HI deficiency as the difference
between the observed HI mass and that expected in an isolated galaxy,  \citet{1985ApJ...292..404G} have 
in fact shown that the HI deficiency increses towards the cluster center. This supports the idea 
that the HI deficiency can trace environmental effects and that galaxies in the same cluster
which are not HI deficient can be considered similar to isolated galaxies. 

Another open question is whether dust-to-gas ratio can be used as metallicity indicator.
If the production of oxygen and carbon are the same and the production, growth and destruction of
solid grains does not change from a galaxy to another, we expect the
fraction of condensed elements in grains to be proportional to those in the gas phase.
Radial analysis of individual galaxies has shown that this is indeed the case
\citep{2011A&A...535A..13M,2010MNRAS.402.1409B} and
models have been proposed to predict how the dust abundance varies as
galaxies evolve in metallicity \citep{1998ApJ...496..145L,2002A&A...388..439H}.
\citet{2007ApJ...663..866D} found  that the dust-to-gas ratio in the central regions does not 
depend on the galaxy morphological type for galaxies of type Sa to Sd. 
But a linear scaling between the metallicity and the dust-to-gas ratio is not always found  \citep{2002A&A...384...33B},
especially when metal abundances are low \citep{2009ApJ...701.1965M}.
However, as shown by \citet{2011A&A...532A..56G}, this might result from
an inaccurate determination of dust mass due to the lack of data coverage at long infrared 
wavelengths. Using the wide  wavelength coverage of {\it Herschel} we investigate whether the 
dust-to-gas ratio can be used as metallicity indicator for metal rich spirals in Virgo. 

We describe the galaxy sample and the database in Section~2,
the relation of the CO brightness and gas masses with the far-infrared emission 
and dust masses in Section~3.  An analysis of possible environmental effects on the dust and  
gas content, atomic and molecular, is given in Section~4. In Section~5 we summarize the main results and 
conclude.  
 
\section{The sample and the data}

Our sample is a magnitude-limited sample of galaxies in the Virgo cluster area mapped by HeViCS. The 
HeViCS covers $\approx 84$ sq. deg. of the Virgo cluster, not the whole cluster area \citep{2010A&A...518L..48D},
and the full-depth region observed with both PACS and SPIRE is limited to 55 sq. deg., due to offsets between scans.
Galaxies in the Virgo Cluster Catalogue (VCC) \citep{1985AJ.....90.1681B} are included in our sample only 
if they are of morphological type Sab to Sm, are located in the HeViCS fields, their distance  
is $17\le D \le 32$~Mpc, and their total magnitude in the B-system, as given in the Third Reference Catalog 
of Bright Galaxies \citep{1995yCat.7155....0D}, is BT$< 13.04$.   
This limiting magnitude ensures that all galaxies in our sample have been observed and detected at 21-cm and 
in the CO J=1-0 line. 
Very few Virgo galaxies fainter than BT$=13.04$  have been detected in the CO J=1-0 line. 
In the HeViCS area, and for the selected range of morphological type, our sample is 100$\%$ complete down to 
B-mag 13.04. With respect to the 250~$\mu$m flux our sample is 100$\%$ complete down to F$_{250}=$5~Jy. 
The significance of correlations between different galaxy properties are assessed using
the Pearson linear correlation coefficient, $r$. 

The complete  list of 35 galaxies selected in our sample is given in Table~1.
In Column~1  we give the sample code, which indicates the type of CO observations
available for estimating the molecular mass. "M" refers to the main sample, which consists of all 
galaxies in HeViCS  that have been fully mapped in the CO J=1-0 line. "S" refers to the 
secondary sample: galaxies in HeViCS which have been detected but not  fully mapped in CO. 
The galaxy name, right ascension and declination are given in Columns 2, 3, and 4 respectively. 
The distance to the galaxy in Mpc, the galaxy morphological type and stellar
mass are given in Columns 5, 6, and 7, respectively. Distances are determined from the average 
redshift of the individual cluster sub-groups \citep{1999MNRAS.304..595G} and we adopts the morphological    
classification scheme of the VCC. Nebular oxygen abundances, 12+log(O/H), from the Sloan Sky
Digital Survey are shown in Column 8. In parenthesis  the metallicity inferred from the
stellar mass using the mass-metallicity relation of 
\citet{2004ApJ...613..898T} and adopting the functional form fit of \citet{2008ApJ...681.1183K}.
The atomic and molecular hydrogen mass are  listed in Columns 9, 10-11, respectively. 
The two values of the molecular  mass, M$^c_{H_2}$ and M$^v_{H_2}$, have been derived from the 
integrated CO brightness using two different values of the
CO-to-H$_2$ conversion factor: a uniform constant value of X$_{CO}$  
or a metallicity dependent CO-to-H$_2$ conversion factor (as given in more details in Section~2.1).  In 
Column~12 we give the relevant reference for the integrated CO line intensity. In the last column
we indicate the number of positions where CO has been detected/observed or whether the galaxy has been fully 
mapped. If more than one map is available we take the mean CO brightness and list 
the references joined by a plus sign. 

We derive the stellar mass of each galaxy, M$_*$, using
the B-K color and the K-band luminosity, as described in \citet{2005A&A...433..807M}. In order to have a 
uniform set of photometric data, we use the GOLDMINE database\footnote{http:http://goldmine.mib.infn.it/}
\citep{2003A&A...400..451G} for the H$\alpha$, B-, H- and K-band magnitudes  
\citep{1996ApL&C..35....1G}, and for the disk geometrical parameters \citep{1973ugcg.book.....N}. 
We derive absolute luminosities after correcting for internal extinction according to 
the prescription given by \citet{1996ApL&C..35....1G}. Corrections for foreground galactic
extinction are small ($A_V\simeq 0.07$). 

As a metallicity indicator we use  the O/H abundance obtained by
\citet{2004ApJ...613..898T} from the SDSS data. Metallicity has been estimated by  fitting
at the same time  all prominent emission lines ([O II], H$\beta$, [O III], H$\alpha$, [N
II], [S II]) with an integrated galaxy spectrum model \citep{2001MNRAS.323..887C}. Not all  
galaxies in our sample have SDSS spectra with measurable emission lines or appropriate line
ratios needed to derive O/H abundances. This is because SDSS narrow slits are generally positioned on the  
central high surface brightness regions  and the presence of some nuclear activity or of a bulge, prevents 
the authors  from probing emission line HII regions in the disk. Galaxies with very weak emission
lines or with anomalous line ratios because of nuclear activity have been discarded by 
\citet{2004ApJ...613..898T}. Thanks to the analysis of large galaxy samples 
from the SDSS, the  mass-metallicity relation predicts 
that metallicity scales with stellar mass until
the stellar mass is below $\sim 3\times 10^{10}$~M$_\odot$, then it flattens out.  
Our sample is representative of the mass-metallicity relation:
the dispersion of the original dataset around the functional 
form fit is $\approx 0.15$~dex \citep{2004ApJ...613..898T,2008ApJ...681.1183K}, 
which is consistent with that of our sample. In this paper we shall use the O/H abundance derived from the 
mass-metallicity relation for all galaxies with known K-band luminosity. 

Values of atomic hydrogen  mass are taken from the VLA Imaging of Virgo Spirals in Atomic Gas
database  \citep[VIVA,][]{2009AJ....138.1741C}, where possible. We correct the HI masses to reflect
the galaxy distances used in the present paper if they differ from the value of 17~Mpc adopted in
the VIVA survey.  \citet{2009AJ....138.1741C} have found an excellent agreement
between single dish fluxes and VIVA fluxes, with no indication that  VIVA data
is missing any very extended flux. 
If the galaxy is not in VIVA the GOLDMINE value of the HI mass is used. 
We will use M$_{gas}$ to indicate the total hydrogen mass of a galaxy i.e. the
sum of the atomic (M$_{HI}$) and molecular hydrogen mass (M$_{H_2}$)  
corrected for helium abundance.

\begin{table*}
\caption{Galaxies in the M and S samples and related quantities. }
\label{sampletab}
\begin{minipage}{\textwidth}
\begin{center}
\begin{tabular}{lcccccccccccc}
\hline \hline
       & ID & RA  & DEC & D & T& M$_*$ & O/H &  log M$_{HI}$&log M$^c_{H_2}$&log M$^v_{H_2}$& Ref. & Notes \\
       &    & deg & deg &Mpc& &M$_\odot$& dex   & M$_\odot$ & M$_\odot$ & M$_\odot$ & \\
(1)    & (2)& (3) & (4) &(5)&(6)&(7)& (8)&  (9)      & (10)      & (11)& (12) & (13)  \\
\hline \hline

M & NGC 4189& 183.447& 13.425& 32& Sc  &12.69 &9.1(9.1)      &9.30$^a$ &9.26 &9.37    & 1 (2)& map\\
M & NGC 4192& 183.451& 14.900& 17& Sb  &10.73 &(..) (9.1)    &9.63$^a$ &9.05 &8.90    & 3 &   map\\
M & NGC 4212& 183.914& 13.901& 17& Sc  &11.82 &9.1(9.1)      &8.91     &9.05 &8.96    & 3 &   map\\
M & NGC 4254& 184.707& 14.416& 17& Sc  &10.45 &9.2(9.1)      &9.65$^a$ &9.88 &9.60    & 4+3 & map\\
M & NGC 4298& 185.387& 14.606& 17& Sc  &11.95 &9.1(9.1)      &8.69$^a$ &9.02 &8.96    & 4+1 & map\\
M & NGC 4302& 185.427& 14.598& 17& Sc  &12.31 &(..) (9.1)    &9.17$^a$ &9.09 &8.93    & 4 &   map\\
M & NGC 4303& 185.479&  4.474& 17& Sc  &10.30 &9.2(9.1)      &9.68     &9.76 &9.49    & 4+3 & map\\
M & NGC 4321& 185.729& 15.823& 17& Sc  &10.02 &9.2(9.1)      &9.46$^a$ &9.79 &9.39    & 4+3 & map\\
M & NGC 4388& 186.445& 12.662& 17& Sab &11.87 &8.8(9.1)      &8.57$^a$ &8.85 &8.90    & 1 (5,2) & map\\
M & NGC 4402& 186.531& 13.113& 17& Sc  &12.64 &9.0(9.1)      &8.57$^a$ &9.16 &9.10    & 4+3 & map\\
M & NGC 4438& 186.940& 13.009& 17& Sb  &11.12 &(..) (9.1)    &8.68     &8.83 &8.56    & 4 &   map\\
M & NGC 4501& 187.997& 14.421& 17& Sbc &10.50 &8.8(9.1)      &9.22$^a$ &9.75 &9.90    & 4+3 & map\\
M & NGC 4522& 188.415&  9.175& 17& Sbc &12.97 &9.0(9.0)      &8.53$^a$ &8.48 &8.51    & 5 &   map\\
M & NGC 4535& 188.585&  8.198& 17& Sc  &10.73 &(..) (9.1)    &9.52$^a$ &9.50 &9.25    & 4+3 & map\\
M & NGC 4567& 189.136& 11.258& 17& Sc  &11.91 &(..) (9.1)    &8.97$^a$ &9.05 &8.97    & 4 &   map\\
M & NGC 4568& 189.143& 11.239& 17& Sc  &11.22 &9.2(9.1)      &9.18$^a$ &9.35 &9.03    & 4 &   map\\
M & NGC 4569& 189.208& 13.163& 17& Sab &10.08 &(..) (9.1)    &8.79$^a$ &9.46 &9.16    & 4+3 & map\\
M & NGC 4579& 189.431& 11.818& 17& Sab &10.52 &(..) (9.1)    &8.75$^a$ &9.18 &9.04    & 4+3 & map\\
S & NGC 4152& 182.656& 16.033& 32& Sc  &12.77 &9.3(9.1)      &9.73     &9.20 &9.13    & 6 & 1/1\\ 
S & NGC 4206& 183.820& 13.024& 17& Sc  &13.00 &(..) (9.0)    &9.38     &8.47$^{b}$ &8.50$^{b}$    & 6 & 2/5\\
S & NGC 4216& 183.977& 13.149& 17& Sb  &10.84 &(..) (9.1)    &9.25$^a$ &9.10 &9.03    & 2 & 4/9\\
S & NGC 4237& 184.298& 15.324& 17& Sc  &12.66 &(..) (9.1)    &8.32     &9.15 &9.08    & 6 & 1/5\\
S & NGC 4273& 184.983&  5.343& 32& Sc  &12.49 &9.1(9.1)      &9.54     &9.32 &9.25    & 2 & 2/3\\
S & NGC 4294& 185.324& 11.511& 17& Sc  &12.68 &(..) (9.0)    &9.21$^a$ &7.85 &7.88    & 1 (2) & 2/2\\
S & NGC 4299& 185.420& 11.503& 17& Scd &12.99 &8.8(8.6)      &9.04$^a$ &8.08 &8.53    & 1 (2) & map\\
S & NGC 4307& 185.523&  9.044& 23& Sbc &12.78 &(..) (9.1)    &8.15     &8.97 &8.90    & 7 & 3/3\\
S & NGC 4312& 185.630& 15.538& 17& Sab &12.46 &9.1(9.0)      &8.08     &8.87 &8.91    & 2 & 1/3\\
S & NGC 4313& 185.660& 11.801& 17& Sab &12.50 &(..) (9.1)    &8.02     &8.82 &8.75    & 7 & 1/3\\
S & NGC 4351& 186.006& 12.205& 17& Sc  &12.99 &9.0(8.9)      &8.48$^a$ &7.86 &8.00    & 1 (5) & 1/1\\
S & NGC 4380& 186.342& 10.017& 23& Sab &12.43 &(..) (9.1)    &8.37$^a$ &8.76 &8.69    & 8 & 1/1\\
S & NGC 4413& 186.634& 12.611& 17& Sbc &12.94 &9.1(9.0)      &8.29     &8.43 &8.46    & 9 & 1/1\\
S & VCC 939 & 186.697&  8.885& 23& Sc  &13.06 &(..) (9.0)    &9.34     &8.31 &8.35    & 8 & 1/1\\
S & NGC 4430& 186.860&  6.263& 23& Sc  &12.76 &9.0(8.9)      &8.86     &8.76 &8.90    & 9 & 1/1\\
S & NGC 4519& 188.376&  8.654& 17& Sc  &12.46 &9.1(9.0)      &9.43     &8.32 &8.35    & 8 & 1/1\\
S & NGC 4571& 189.235& 14.217& 17& Sc  &12.15 &(..) (9.1)    &8.79     &8.93 &8.91    & 2 & 3/5\\
  
\hline \hline
\end{tabular}
\tablebib{
$(a)$HI mass from the VIVA database; $(b)$this mass is highly uncertain because it has been estimated using a CO 
flux only a factor two above the rms; $(c)$ H$_2$ mass for a constant CO-to-H$_2$ conversion factor;
$(v)$ H$_2$ mass for a  metallicity dependent CO-to-H$_2$ conversion factor. 
References (Column 12).
(1)new IRAM-30mt observations, C.Pappalardo et al. (2012) (in parentheses the reference to the older, 
lower sensitivity published data),
(2)\citet{1995ApJS...98..219Y},(3)\citet{2007PASJ...59..117K},(4)\citet{2009ApJS..184..199C},
(5)\citet{2008A&A...491..455V},(6)\citet{1986ApJ...310..660S},(7)Boselli private communication 
(see also GOLDMINE database),(8)\citet{2002A&A...384...33B},(9)\citet{1995A&AS..110..521B}
} 
\end{center}
\end{minipage}
\end{table*}

\subsection{Integrated CO J=1-0 brightness and molecular mass estimates}

Extensive searches for CO emission in Virgo galaxies have been 
carried out by a number of authors using single dish mm-telescopes. Only 18 galaxies in
the HeViCS area have been mapped (M-sample, plotted with filled circles). 
Some galaxies have been mapped by more than one team using different 
telescopes. In this case the total CO fluxes are in good agreement and  we use the 
arithmetic mean of the two values. Uncertainties from the original papers have been adopted. 
We include in the M-sample NGC4522 which has been mapped in the CO J=2-1 line
and observed at several positions in the CO J=1-0 to determine the 2-1/1-0
line ratio. We do not include NGC4299 whose CO map has a low signal to noise and allows a reliable
determination of the total CO brightness only within the H$\alpha$ emission
map boundary. There are 17 galaxies with BT$< 13.04$ which have been detected in the CO J=1-0 line with one 
or several pointed observations across the optical disk (S-sample, plotted with open circles). 
All but one of these have CO J=1-0 line emission above 3-$\sigma$ in at least one beam.
Only for NGC4206 the CO J=1-0 line has been marginally detected, at the 2-$\sigma$ level. When
the CO line brightness has been measured in more than one position 
we rely on the total CO flux estimated in the original papers. In these cases  the total flux is determined
via exponential or gaussian fits to the radial distribution \citep{1986ApJ...310..660S,1995ApJS...98..219Y}. 
If a galaxy has been detected at just one
position we determine its total CO flux assuming an exponential radial distribution, centrally peaked, with
scalelength $\lambda=0.2 R_{25}$.

The value $\lambda=0.2\pm0.1 R_{25}$ is the average radial scalelength of the CO flux distribution 
inferred for the M-sample (mapped galaxies). This is consistent with the exponential
scalelengths measured for the CO emission in nearby galaxies mapped in the Heterodyne Receiver Array 
CO Line Extragalactic Survey \citep{2009AJ....137.4670L}. 
We convolve the exponential function, corrected for disk inclination,
with the telescope beam, and compare this to the observed flux in order to determine the central brightness
and hence the total CO flux.   
Notice that  S-sample galaxies, as well as mapped Virgo galaxies, do not have enhanced CO emission in the center 
and are very well represented by exponential functions, as shown by   \citet{2007PASJ...59..117K}. For
HeViCS galaxies in the \citet{2007PASJ...59..117K} sample the fraction of molecular gas in the inner regions, 
f$_{in}$, is only 0.14 on average, for both Seyfert and non-Seyfert galaxies. 
The dispersion between the total CO flux  as measured from the map,  and the value derived from the exponential fit
with scalelength $\lambda=0.2 R_{25}$ is 0.15 in log units. This is the error which we add in quadrature to the 
flux measurement error when computing the uncertainties in the total CO brightness of galaxies in the S-sample. 

For 6 galaxies in Table 1 we use newly acquired data from observations  with the
IRAM 30-m telescope (Pappalardo et al. 2012). Here we give a  brief description of the observations.
We  searched for CO J=1-0 and J=2-1 lines in NGC4189, NGC4294, NGC4298,
NGC4299, NGC4351, NGC4388, during June 2010 and April 2011. The FWHM beam is 22~arcsec at 115~GHz, 
the frequency of the J=1-0 line used in this paper.  
We observed the sources in position-switching mode, using the EMIR receiver combination E0/E2  
and the VESPA and WILMA backend system with a bandwidth of 480~MHz and 4~GHz, respectively. 
The on-the-fly mapping mode was used for mapped galaxies. Non-mapped
galaxies were sampled along the major axis at 22~arcsec spacing.
The spectra were smoothed in velocity to 10.5~km~s$^{-1}$ and the data from different backends were averaged resulting 
in an rms between 7 and 19~mK. 
 
Two values of the molecular hydrogen mass for each galaxy are listed in Column 10 and 11 of Table 1. 
In Column 10 $M_{H_2}^c$ is derived 
from the total integrated CO flux using the given distance and a constant CO-to-H$_2$ conversion factor, equal
to that found in the solar neighborhood  X$_{CO}\simeq 2\times 10^{20}$~mols.~cm$^{-2}$/(K~km~s$^{-1}$) 
\citep{1996A&A...308L..21S,2001ApJ...547..792D,2010ApJ...710..133A,2011MNRAS.415.3253S}.
In Column 11 $M_{H_2}^v$ is the molecular hydrogen mass for a  metallicity dependent CO-to-H$_2$ conversion factor
computed as follows. We assume that the metallicity at 
the galaxy center, Z$_c$, is that given by the mass-metallicity relation (only for VCC975 we do take the SDSS value  
because of the unknown K-band magnitude) and that there is a radial metallicity gradient similar to that 
determined by the oxygen abundances in our own galaxy through optical studies \citep{2006ApJS..162..346R}.
This can be rewritten as:

\begin{equation}
Z = -0.8 {R\over R_{25}} + Z_c
\end{equation}  

\noindent
where Z=12+log O/H.
Using the above equation and the value of the CO-to-H$_2$ conversion factor measured in the solar neighborhood 
(at $R\simeq 0.67 R_{25}$) the following radial dependence can be written:
 
\begin{equation}
{\hbox{log}} {X^v_{CO}\over 2\times 10^{20}} = 0.8 {R\over R_{25}} - Z_c + 8.67
\end{equation} 

\noindent
if one assumes a CO-to-H$_2$ conversion factor which varies inversely with metallicity.
For $Z_c=9.1$, a typical value of the central metallicity in our sample, the above equation predicts that the
conversion factor is 8$\times 10^{19}$~cm$^{-2}$/(K~km~s$^{-1}$) at the center and 
5$\times 10^{20}$~cm$^{-2}$/(K~km~s$^{-1}$) at $R=R_{25}$.  Unless stated differently we shall use $X^v_{CO}$ and the 
corresponding masses (Column 11 in Table 1) throughout this paper. However, due to the lack of data for
radial metallicity gradients in cluster galaxies  and to uncertainties of metallicity calibrators, we
will quote also results relative to $M_{H_2}^c$. Given the limited metallicity range of our sample
results are not much dependent on the CO-to-H$_2$ factor used  
 \citep{2008ApJ...686..948B,2011ApJ...737...12L}.

\subsection{  Far-infrared observations and dust mass estimates} 

For the far-infrared (hereafter FIR)  emission  we use the HeViCS dataset described in \citet{2012MNRAS.419.3505D}
but with all the eight scans combined, as described in Auld et al. (2012), since  
Herschel observations have now been completed. The Herschel satellites has scanned the whole area of the survey 
four times in two orthogonal directions giving a total of eight images for each band.  The full 
width half maximum (FWHM) beam sizes are approximately 9.4 
and 13.5~arcsec with pixel sizes of 3.2 and 6.4~arcsec for the 100 and 160~$\mu$m
PACS channels, respectively. The FWHM of the SPIRE beams are 18.1, 25.2, and 36.9~arcsec
with pixel sizes of 6, 8, and 12~arcsec at 250, 350, and 500~$\mu$m, respectively.
  
As shown in \citet{2010A&A...518L..65B}, emission longward of 100~$\mu$m traces
the cold dust heated by evolved stars which constitutes the bulk of dust mass in spiral galaxies.
We can derive dust masses and temperatures using the PACS and SPIRE
fluxes after fitting the spectral energy distribution (SED) between 100 and 500~$\mu$m.
In Table 2 we show the Herschel fluxes, dust masses and temperatures, and the $\chi^2$ values for the SED fits 
relative to all galaxies in our 
sample. Dots indicate that the galaxy is outside the area mapped by HeViCS in that band.   
The errors quoted in Table~2 cover uncertainties due to the choice of the aperture, to the calibration 
and to  photometric random errors (Auld et al. 2012). Calibration uncertainties are the largest errors also 
for PACS photometry, (in this case they have been assumed to be of order of 20$\%$). For bright galaxies,
such as those in our sample, the procedure and dataset described in Auld et al. (2012) gives 
fluxes, dust temperatures and masses which are consistent with those obtained by \citet{2012MNRAS.419.3505D}
using only two orthogonal scans i.e. one quarter of the final exposure times.   
Using the complete HeViCS dataset we detect all galaxies in all 5 bands. For only one galaxy, NGC 4571, 
we cannot measure PACS fluxes because it lies outside the boundaries of the PACS mapped area.

\begin{table*}
\caption{Herschel fluxes in the various bands, dust masses, temperatures and reduced $\chi^2$ for the SED fit.}
\label{firtab}
\begin{minipage}{\textwidth}
\begin{center}
\begin{tabular}{lcccccccccc}
\hline \hline
Sample& ID &  F$_{500}$ & F$_{350}$ & F$_{250}$ & F$_{160}$ & F$_{100}$ & log M$_{dust}$ & T$_{dust}$ & $\chi^2$     \\
      &     &   Jy       &   Jy      &    Jy     &    Jy     &    Jy    & M$_\odot$   &          &               \\
 (1)  &(2)  &(3)        &  (4)      & (5)       & (6)       &   (7)     & (8)         & (9)      &(10)          \\       

\hline \hline
M & NGC 4189&  1.02$\pm$0.08&  2.99$\pm$0.22&   7.05$\pm$0.50&  11.43$\pm$2.29&  10.10$\pm$2.02& 7.6$\pm$0.1& 22.0$\pm$0.9& 0.59 \\
M & NGC 4192&  5.03$\pm$0.37& 13.14$\pm$0.95&  28.45$\pm$2.02&  34.80$\pm$6.96&  22.99$\pm$4.60& 7.9$\pm$0.1& 18.7$\pm$0.7& 0.68 \\
M & NGC 4212&  1.88$\pm$0.14&  5.43$\pm$0.39&  13.38$\pm$0.94&  22.79$\pm$4.56&  19.85$\pm$3.97& 7.3$\pm$0.1& 22.4$\pm$0.9& 0.88 \\
M & NGC 4254&  9.28$\pm$0.66& 27.67$\pm$1.96&  68.36$\pm$4.80&  117.3$\pm$23.5&  100.3$\pm$20.0& 8.0$\pm$0.1& 22.8$\pm$0.9& 1.36 \\
M & NGC 4298&  1.87$\pm$0.14&  5.30$\pm$0.37&  12.56$\pm$0.88&  19.56$\pm$3.91&  13.98$\pm$2.80& 7.4$\pm$0.1& 21.0$\pm$0.8& 1.20 \\ 
M & NGC 4302&  3.23$\pm$0.23&  8.77$\pm$0.63&  19.84$\pm$1.41&  25.81$\pm$5.16&  16.63$\pm$3.33& 7.7$\pm$0.1& 19.5$\pm$0.7& 1.28 \\
M & NGC 4303&  8.27$\pm$0.59& 23.51$\pm$1.66&  56.24$\pm$3.95&  95.96$\pm$19.2&  90.52$\pm$18.1& 7.9$\pm$0.1& 22.4$\pm$1.0& 0.29 \\
M & NGC 4321& 10.37$\pm$0.74& 29.68$\pm$2.10&  70.70$\pm$4.96&  94.58$\pm$18.9&  70.32$\pm$14.1& 8.1$\pm$0.1& 20.6$\pm$0.7& 2.28 \\  
M & NGC 4388&  1.38$\pm$0.10&  3.67$\pm$0.26&   9.31$\pm$0.66&  17.37$\pm$3.47&  18.46$\pm$3.69& 7.1$\pm$0.1& 23.4$\pm$1.2& 0.84 \\
M & NGC 4402&  2.22$\pm$0.16&  6.27$\pm$0.44&  14.97$\pm$1.05&  24.10$\pm$4.82&  18.45$\pm$3.69& 7.4$\pm$0.1& 21.3$\pm$0.8& 0.86 \\
M & NGC 4438&  1.40$\pm$0.13&  3.81$\pm$0.29&   8.71$\pm$0.63&  12.33$\pm$2.47&  10.65$\pm$2.13& 7.3$\pm$0.1& 20.5$\pm$0.9& 0.51 \\
M & NGC 4501&  9.24$\pm$0.65& 25.19$\pm$1.78&  61.21$\pm$4.30&  94.14$\pm$18.8&  72.34$\pm$14.5& 8.0$\pm$0.1& 21.2$\pm$0.8& 1.09 \\
M & NGC 4522&  0.59$\pm$0.05&  1.57$\pm$0.12&   3.40$\pm$0.25&   5.12$\pm$1.03&   4.13$\pm$0.83& 6.9$\pm$0.1& 20.2$\pm$0.9& 0.17 \\ 
M & NGC 4535&  6.33$\pm$0.48& 16.27$\pm$1.17&  35.37$\pm$2.49&  41.18$\pm$8.24&  23.76$\pm$4.76& 8.0$\pm$0.1& 18.4$\pm$0.6& 1.13 \\
M & NGC 4567&  1.70$\pm$0.30&  4.70$\pm$0.90&  10.80$\pm$2.20&  20.50$\pm$4.10&  16.10$\pm$3.20& 7.3$\pm$0.1& 21.8$\pm$1.0& 0.22 \\       
M & NGC 4568&  4.20$\pm$0.84& 12.00$\pm$2.40&  30.80$\pm$6.20&  51.10$\pm$10.2&  46.10$\pm$9.20& 7.6$\pm$0.1& 22.9$\pm$0.9& 0.12 \\    
M & NGC 4569&  3.27$\pm$0.24&  9.18$\pm$0.65&  22.97$\pm$1.62&  35.95$\pm$7.19&  28.35$\pm$5.67& 7.6$\pm$0.1& 21.6$\pm$0.7& 1.64 \\
M & NGC 4579&  3.36$\pm$0.25&  9.40$\pm$0.69&  21.79$\pm$1.57&  30.80$\pm$6.16&  20.38$\pm$4.08& 7.6$\pm$0.1& 20.2$\pm$0.7& 1.42 \\ 
S & NGC 4152&  0.81$\pm$0.07&  2.18$\pm$0.16&   5.31$\pm$0.38&   9.91$\pm$1.98&   9.72$\pm$1.95& 7.5$\pm$0.1& 22.8$\pm$1.2& 0.36 \\
S & NGC 4206&  1.08$\pm$0.09&  2.25$\pm$0.17&   3.89$\pm$0.28&   4.08$\pm$0.82&   2.22$\pm$0.45& 7.3$\pm$0.1& 15.6$\pm$0.8& 2.73 \\
S & NGC 4216&  4.01$\pm$0.29& 10.41$\pm$0.74&  21.92$\pm$1.54&  27.06$\pm$5.41&  15.16$\pm$3.04& 7.8$\pm$0.1& 18.2$\pm$0.6& 0.56 \\
S & NGC 4237&  1.12$\pm$0.09&  3.21$\pm$0.23&   7.88$\pm$0.56&  12.78$\pm$2.56&   9.84$\pm$1.97& 7.1$\pm$0.1& 21.7$\pm$0.9& 1.34 \\ 
S & NGC 4273&  1.56$\pm$0.12&  4.24$\pm$0.30&  10.54$\pm$0.74&  21.01$\pm$4.20&  21.76$\pm$4.35& 7.7$\pm$0.1& 23.7$\pm$1.3& 0.36 \\
S & NGC 4294&  0.90$\pm$0.07&  2.12$\pm$0.16&   4.23$\pm$0.30&   6.87$\pm$1.38&   5.99$\pm$1.20& 7.1$\pm$0.1& 19.0$\pm$1.1& 2.79 \\
S & NGC 4299&  0.49$\pm$0.05&  1.21$\pm$0.10&   2.58$\pm$0.19&   4.45$\pm$0.89&   4.95$\pm$0.99& 6.7$\pm$0.1& 20.9$\pm$1.5& 2.15 \\ 
S & NGC 4307&  0.68$\pm$0.05&  1.90$\pm$0.14&   4.41$\pm$0.31&   6.34$\pm$1.27&   4.42$\pm$0.89& 7.2$\pm$0.1& 20.3$\pm$0.8& 1.14 \\
S & NGC 4312&  0.53$\pm$0.04&  1.65$\pm$0.12&   4.20$\pm$0.30&   7.64$\pm$1.53&   6.36$\pm$1.27& 6.8$\pm$0.1& 23.2$\pm$1.0& 2.24 \\
S & NGC 4313&  0.60$\pm$0.05&  1.72$\pm$0.13&   4.15$\pm$0.30&   6.13$\pm$1.23&   4.37$\pm$0.88& 6.9$\pm$0.1& 20.9$\pm$0.8& 1.69 \\
S & NGC 4351&  0.28$\pm$0.03&  0.77$\pm$0.07&   1.64$\pm$0.12&   2.48$\pm$0.50&   1.74$\pm$0.35& 6.6$\pm$0.1& 19.9$\pm$0.9& 0.07 \\
S & NGC 4380&  0.80$\pm$0.07&  2.21$\pm$0.16&   4.89$\pm$0.35&   6.12$\pm$1.22&   2.96$\pm$0.60& 7.4$\pm$0.1& 18.5$\pm$0.5& 2.57 \\
S & NGC 4413&  0.44$\pm$0.04&  1.15$\pm$0.09&   2.55$\pm$0.19&   3.98$\pm$0.80&   2.81$\pm$0.56& 6.8$\pm$0.1& 20.0$\pm$0.9& 0.11 \\
S & VCC 939 &  0.61$\pm$0.06&  1.25$\pm$0.12&   2.23$\pm$0.18&   2.03$\pm$0.41&   0.63$\pm$0.14& 7.4$\pm$0.1& 14.9$\pm$0.5& 0.52 \\
S & NGC 4430&  0.67$\pm$0.06&  1.83$\pm$0.14&   4.06$\pm$0.29&   5.97$\pm$1.20&   3.90$\pm$0.78& 7.2$\pm$0.1& 19.8$\pm$0.8& 0.39 \\
S & NGC 4519&  1.03$\pm$0.09&  2.54$\pm$0.19&   5.24$\pm$0.37&   8.12$\pm$1.63&   6.30$\pm$1.26& 7.1$\pm$0.1& 19.4$\pm$0.9& 0.87 \\
S & NGC 4571&  1.36$\pm$0.12&  4.00$\pm$0.31&   8.92$\pm$0.68&         (...)  &          (...) & 7.1$\pm$0.1& 22.7$\pm$3.6& 0.01 \\

\hline \hline
\end{tabular}
\end{center}
\end{minipage}
\end{table*}

The SED fits were done for each galaxy after convolving the modified blackbody model with the spectral response function 
for each band, thus automatically taking into account color corrections for the specific SED of each
object. We used a power law dust emissivity $\kappa_\lambda=\kappa_0(\lambda_0/\lambda)^\beta$, with spectral
index $\beta=2$ and emissivity $\kappa_0=0.192$~m$^2$~kg$^{-1}$ at $\lambda_0$=350$\mu$m. These values reproduce 
the average emissivity of models of the Milky Way dust in the FIR-submm \citep{2003ARA&A..41..241D}. They are
also consistent with the dust spectral emissivity index $\beta$ determined in molecular clouds
using the recent {\it Planck} data \citep[e.g.][and references therein]{2011A&A...536A..25P}.  
The standard $\chi^2$  minimization technique is used in order to derive the dust temperature and mass. 
Uncertainties in the temperature determination, considering the photometry errors, are about 1-2~K.
Uncertainties on dust temperature and mass are only slightly larger for the NGC 4571 for which there is no PACS data.
More details on SED fitting are given in \citet{2010A&A...518L..51S,2011A&A...535A..13M,2012MNRAS.419.3505D}.
Note that single black body SED models fit well S-sample galaxies as well as M-sample galaxies. All SED fits are   
displayed by \citet{2012MNRAS.419.3505D} and by Auld et al. (2012). A two component fit is not well constrained by 
just five data points in the SED and it would require shorter and longer wavelength data.

\begin{figure*}[ht!]
\includegraphics[width=14cm]{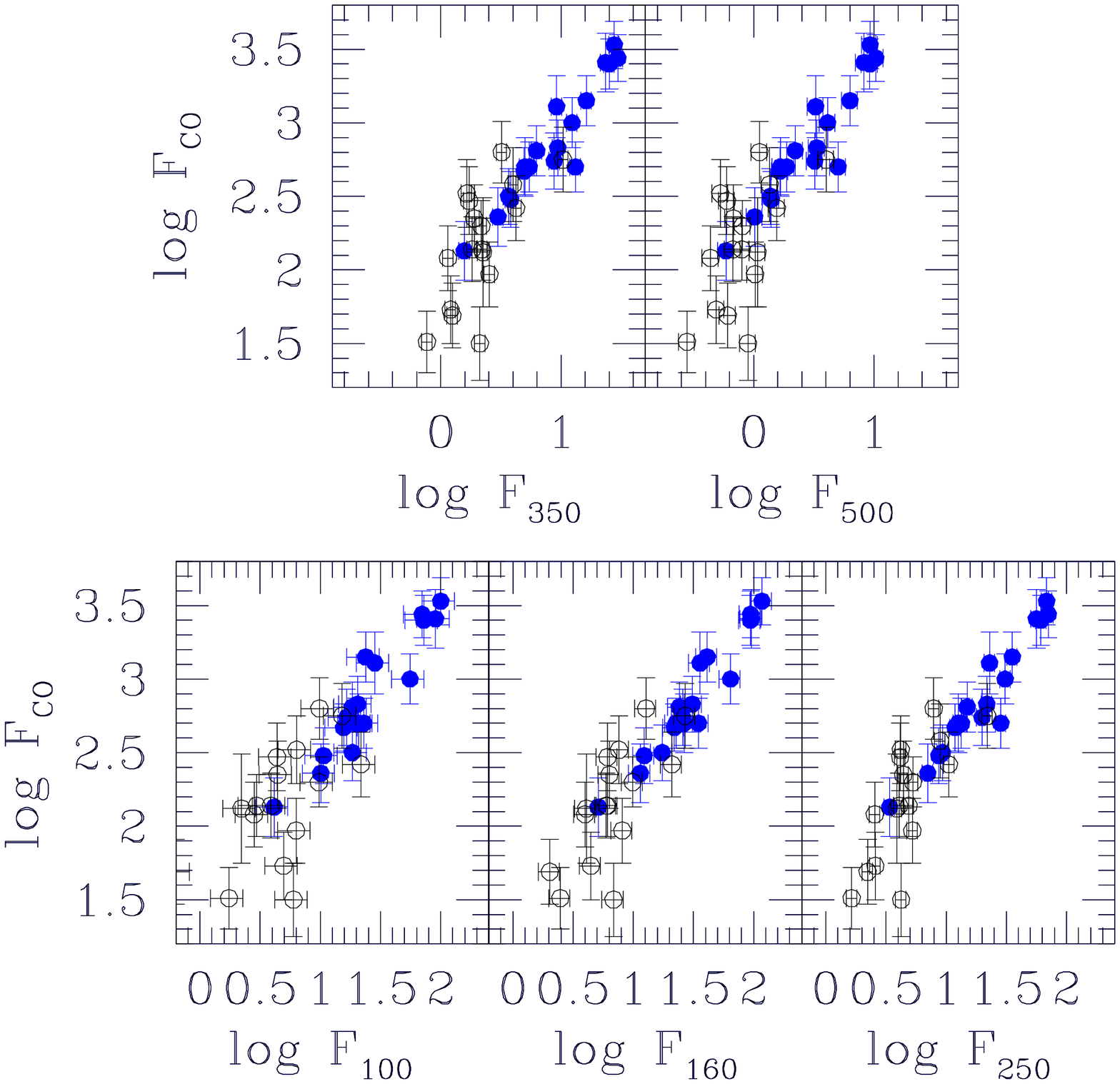}
\caption{Fluxes in the 5 Herschel bands (in Jy  units)
versus the total CO J=1-0 flux (in Jy~km~s$^{-1}$). 
Filled  circles are galaxies in the M-sample, open circles galaxies in the S-sample.}
\label{fig:f1}
\end{figure*}

\section{The correlations between dust and the molecular or total gas mass}

Far-infrared fluxes for galaxies in the M-sample correlate well with the total CO flux, as shown in Fig.~\ref{fig:f1}.  
Pearson linear correlation coefficients $r$ in the log F$_{FIR}$-log F$_{CO}$ plane  
are high for fluxes in all 5 Herschel bands, ranging between 0.88 and 0.93 for the whole sample, 
and between 0.94 and 0.96 for galaxies in the M-sample (the highest being for F$_{160}$ and F$_{250}$). 
Slopes are of order unity (0.92$\pm0.08$,1.04$\pm0.08$,1.08$\pm0.08$,1.10$\pm0.07$,1.09$\pm0.05$ for 
FIR wavelengths from 100 to 500~$\mu$m), meaning that the correlation is close to linear. The best linear fit 
in the logF$_{250}$--logF$_{CO}$ plane gives for example logF$_{CO}=1.08(\pm0.08)$logF$_{250} + 
1.45(\pm0.08)$. These slopes are compatible with those derived  considering only  galaxies in the M-sample 
(1.04$\pm0.13$,1.06$\pm0.11$,1.04$\pm0.10$,1.04$\pm0.08$,1.04$\pm0.05$ for FIR wavelengths from 100 to 500~$\mu$m).   
The less tight relation between the CO line intensity and hot dust emission in Virgo spirals  
\citep[e.g.][for IRAS data]{1986ApJ...310..660S} suggests that molecules are more closely related to cold dust
rather than to dust heated by star formation.

\begin{figure}   
\includegraphics[width=\columnwidth]{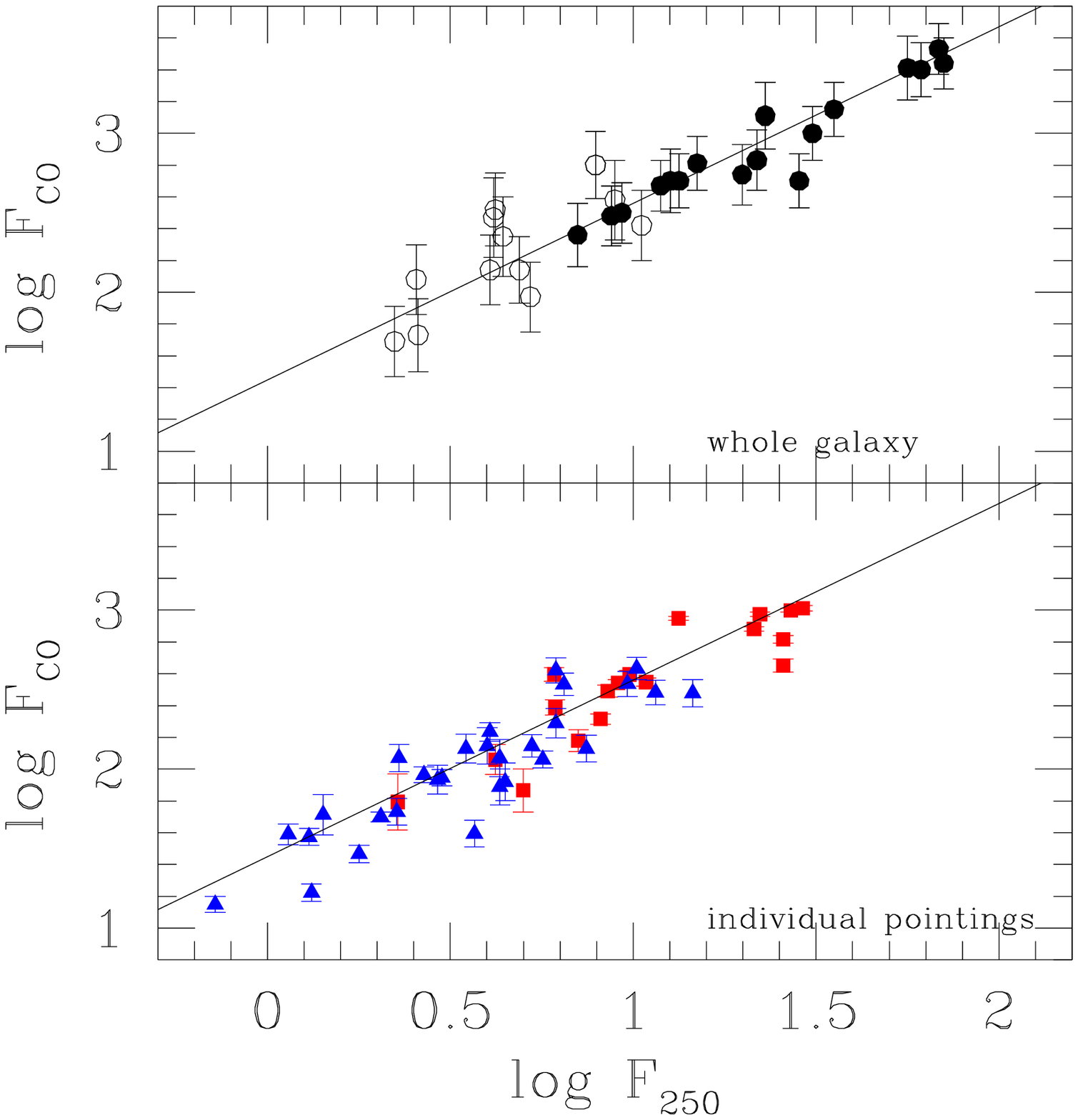}
\caption{{\it Lower panel}: The CO J=1-0 flux, in Jy km s$^{-1}$, measured in one beam at
the galaxy center, versus the 250~$\mu$m flux in Jy  in the same area for galaxies in the M- and S-sample.
The filled squares (red in the on-line version) are for 
beams of 100~arcsec FWHM, while the filled triangles (blue) are for beams  of 45-55~arcsec FWHM. 
{\it Upper panel}: same quantities as in the lower panel but integrated
over the whole galaxy. Filled circles are for galaxies mapped in CO while open circles are for galaxies whose
total integrated CO flux has been inferred using pointed observations (see text for details).}
\label{fig:fn}
\end{figure}

The larger scatter and the slightly lower correlation coefficients found when we include in the analysis galaxies 
in the S-sample, might be driven by the higher uncertainties on the total CO emission, because of the lack of CO maps. 
In order to check whether this is indeed the case, or whether a break in the correlation 
occurs at low luminosities, additional CO maps are needed for galaxies as dim as 13-14~mag in the B-band.
To avoid the uncertainties on the integrated CO fluxes for non mapped galaxies we have checked the slope and 
the scatter of the FIR-CO correlation for galaxy central regions. We convolved the 250~$\mu$m maps from the original
resolution (FWHM=18.2") to the resolution of pointed CO J=1-0 observations (which can be found in references to Table~1).
At the optical center of galaxies we measure the 250~$\mu$m flux  in an area as large as the FWHM of the telescope  
used to measure the CO J=1-0 line intensity (ranging between 45 and 100~arcsec).
Figure~\ref{fig:fn} shows the resulting correlation between F$_{250}$ and the CO flux in the same area, as given in the 
original papers.  The straight line, of slope 1.11$\pm0.02$,
shows the best linear fit to these data. This is compatible with the slope
measured  when values integrated over the galaxy disk are considered (1.04$\pm0.08$).
The  scatter in the correlation is not reduced with respect to that found for the same quantities integrated
over the whole disk.  
 
\begin{figure}[!ht]
\includegraphics[width=\columnwidth]{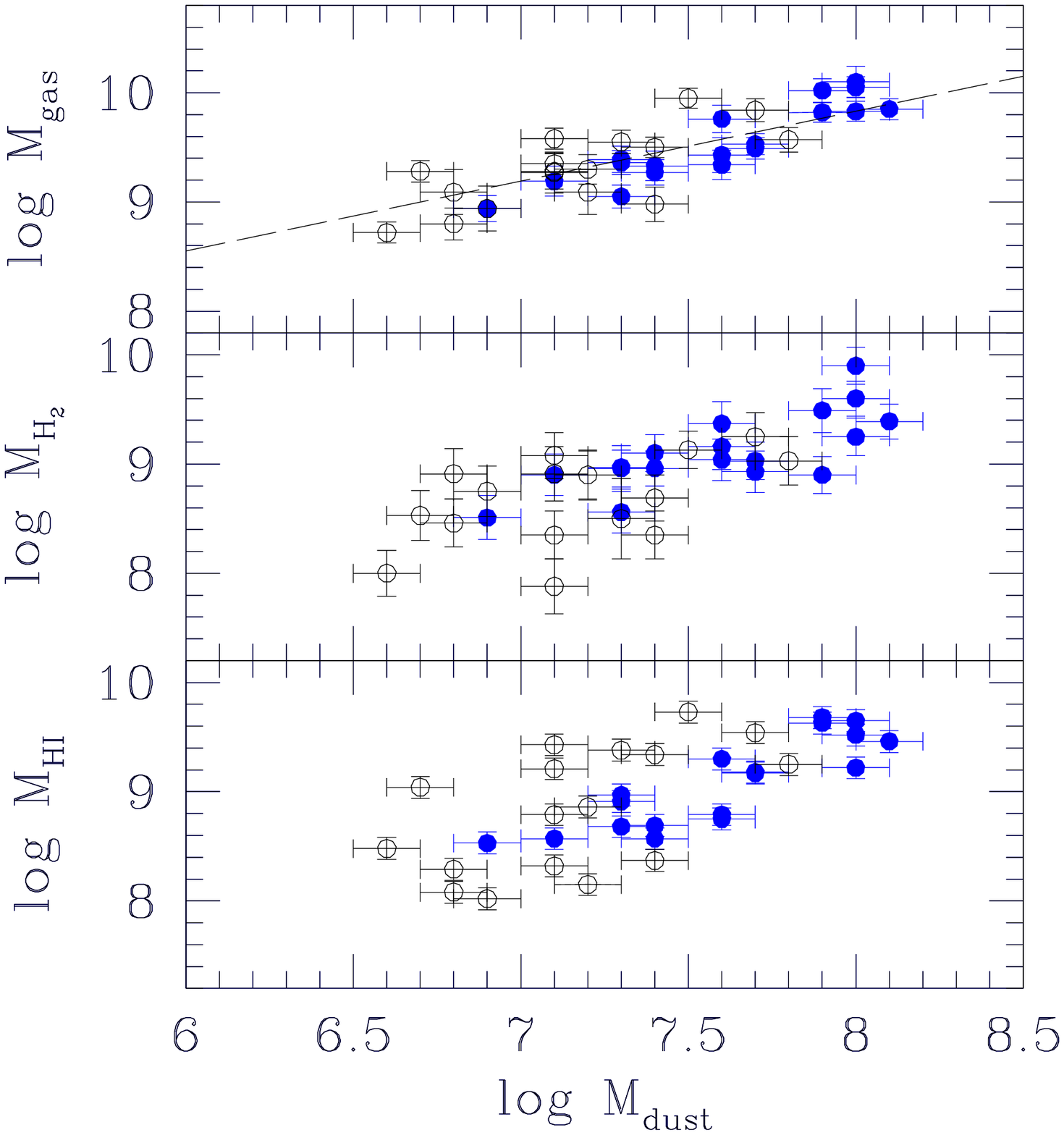}
\caption{The relation between dust and
gaseous masses (HI,H$_2^v$,and total gas mass in solar mass units) for the whole sample.
Filled and open circles  are for galaxies in the M-sample,
and S-sample, respectively. The dashed line in the upper panel is the best-fit
linear relation for the data shown.}
\label{fig:f3}
\end{figure}

Dust masses take into account temperature variations which individual FIR fluxes 
do not. The bluest galaxies have lower dust temperatures i.e a low  F$_{250}$-to-F$_{500}$ flux ratio.
In Figure~\ref{fig:f3} we compare the dust mass, M$_{dust}$, to the molecular,
M$_{H_2}$, atomic, M$_{HI}$, and  total gas mass, M$_{gas}$.
The lowest scatter in the dust-to-molecular gas mass relation is found if M$_{H_2}$=M$_{H_2}^c$,  
for the whole sample ($r=0.81$) or for the M-sample alone ($r=0.88$ in this case). 
The correlation for the whole sample reads:

\begin{equation}
{\hbox{log}} M_{H_2}^c= 0.97(\pm0.13)\times{\hbox{log}} M_{dust} + 1.9(\pm1.0)
\end{equation}

\noindent
A flatter, but still compatible slope  is obtained if one considers the correlation M$_{dust}$--M$_{H_2}^v$:
 
\begin{equation}
{\hbox{log}} M_{H_2}^v= 0.77(\pm0.12)\times{\hbox{log}} M_{dust} + 3.2(\pm0.9)
\end{equation}

\begin{figure}[!ht]
\includegraphics[width=\columnwidth]{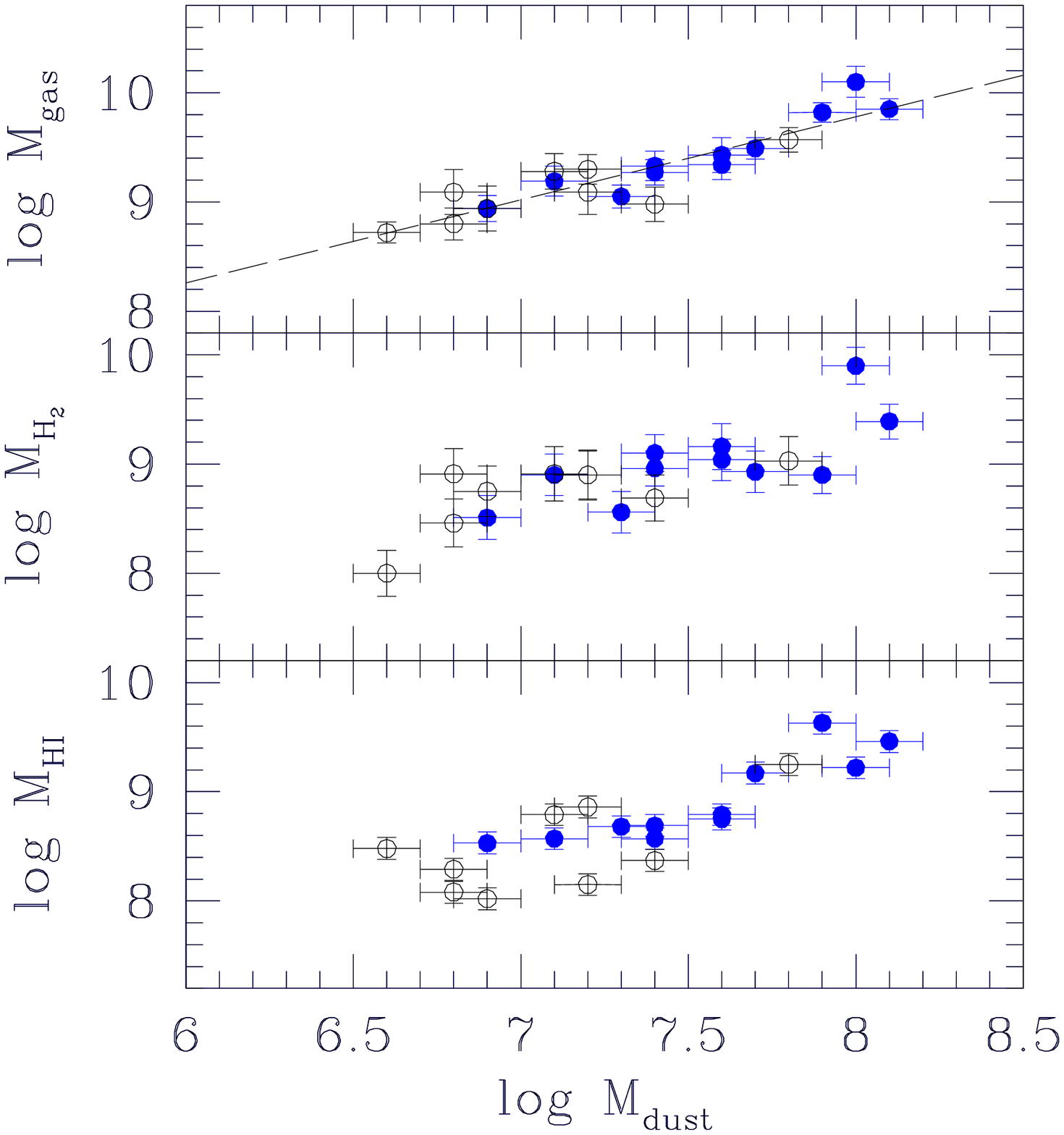}
\caption{The relation between dust and
gaseous masses (in solar mass units) for galaxies with def(HI)$>0.5$. 
Filled and open circles  are for galaxies in the M-sample,
and S-sample, respectively. The dashed line in the upper panel is the best-fit
linear relation for the data shown.}
\label{fig:f23}
\end{figure}

\noindent
The slope is closer to unity if we consider only galaxies of the M-sample, likely because 
this sample is more uniform in galaxy morphology (mostly Sc).
The best correlation with the dust mass is found for the total gas mass, with a Pearson
linear correlation coefficient of 0.91 for the M-sample and of 0.84 for the whole sample.
In the last case the best-fit linear relation between $M_{dust}$ and $M_{gas}$ reads: 

\begin{equation}
{\hbox{log}} M_{gas}= 0.75(\pm0.09)\times{\hbox{log}} M_{dust} + 3.9(\pm0.6)
\end{equation}

\noindent
which implies a lower dust-to-gas ratio for galaxies with a lower gas content. If  M$_{H_2}^c$ 
replaces M$_{H_2}^v$ in the computation of M$_{gas}$ the slope is 0.85$\pm0.08$ and $r=0.88$.
The decrease of the dust-to-gas ratio, as the gas mass decreases,
is not due to the environment: large galaxies, with a large gas mass, are not more HI deficient
than small galaxies (with a small gas mass).  In Figure~\ref{fig:f23} we show the relation between dust 
and gas mass for HI deficient galaxies with def(HI)$>0.5$.  The slope of
the dust-to-total gas mass relation is 0.76$\pm 0.08$, compatible with that found for the whole sample
or for non-HI deficient galaxies (0.62$\pm 0.11$).   As expected,  the dispersion in the M$_{dust}$--M$_{HI}$
relation decreases for HI deficient galaxies, which have dispersed the outer gas in the intercluster
medium (bottom panel of Figure~\ref{fig:f23}). In the mass range covered by our sample, environmental 
effects on the dust-to-gas ratio (see next Section)  do not depend on galaxy mass.
Variations in the dust-to-gas ratio with the global gas content are instead due to internal evolution. 
Smaller galaxies in the S-sample have on average a higher HI content: they have been unable to increase 
the dust abundance and to convert much of their atomic gas into molecular gas
due also to their extended outer HI disks. As shown by Pappalardo et al. (2012) in a spatially 
resolved study, the relation between dust emission and gas surface density in the inner regions is very tight, 
with a slope closer to one for metal rich galaxies.

In the following we convert the flux density in an Herschel band, F$_\nu$, into luminosity 
L$_\nu=4\pi$D$^2\nu$F$_\nu$ and express L$_\nu$ in solar units. Hence we convert the CO J=1-0 total flux in 
Jy~km~s$^{-1}$ into L$_{CO}$ in solar units by multiplying $F_{CO}$ by  0.12 $D_{Mpc}^2$.
The 250~$\mu$m luminosity, similar to M$_{dust}$, correlates better with the
total gas mass rather than with the HI or H$_2$ mass alone.

The scatter in the L$_{CO}$--L$_{B,H,K,H\alpha}$ relation, the B,H,K-band and H$\alpha$ luminosities,
which we corrected for internal extinction, is larger than observed between L$_{CO}$   
and the luminosities in the various Herschel bands. We display L$_{CO}$  
versus L$_{H\alpha}$ and M$_*$ in Figure~\ref{fig:lum} and  
below we give the relations between L$_{CO}$  and L$_B$, L$_{H\alpha}$, M$_*$, L$_{250}$  
for the whole sample (in solar units):

\begin{figure}
\includegraphics[width=\columnwidth]{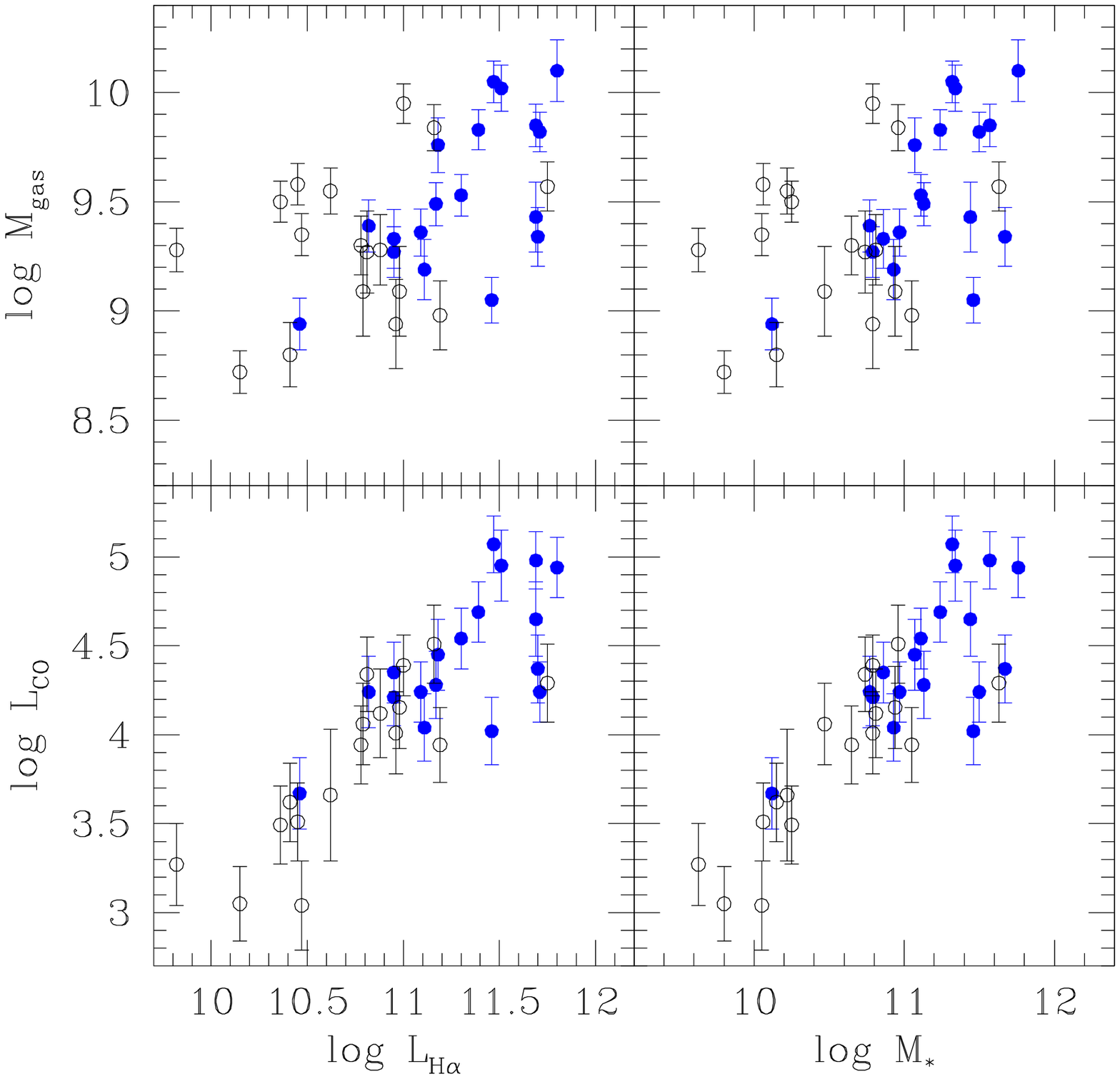}
\caption{The relations between the extinction corrected H$\alpha$ luminosity and stellar mass and the CO
luminosity or total gas mass. Luminosities
and masses are in solar luminosity and  mass units, respectively.  
Symbols for galaxies in different samples are coded as in Figure~\ref{fig:f1}.}
\label{fig:lum}
\end{figure}

\begin{equation}
{\hbox {log L}}_{CO} = 1.10 \times {\hbox{log L}}_B -6.7 \qquad   r = 0.78 
\end{equation}

\begin{equation}
{\hbox {log L}}_{CO} = 0.87 \times {\hbox{log L}}_{H\alpha} -5.4 \qquad   r = 0.83 
\end{equation}

\begin{equation}
{\hbox{log L}}_{CO} = 0.78 \times {\hbox{log M}}_* -4.3 \qquad  r = 0.85 
\end{equation}

\begin{equation}
{\hbox{log L}}_{CO} = 1.07 \times {\hbox{log L}}_{250} -5.6 \qquad  r = 0.92
\label{eq:250h2}
\end{equation}

The correlations of   M$_{gas}$ with L$_{H\alpha}$ or M$_*$  are also less tight (see Figure~\ref{fig:lum}) 
than with L$_{250}$  and much weaker correlations are found for M$_{HI}$.
This confirms that the main correlation with the  CO J=1-0 line intensity is indeed 
established by the cold dust emission. In our sample, which spans a limited metallicity range,
the CO J=1-0 line emission traces molecular hydrogen which 
forms on dust, and this explains the correlation. 
Star formation is not ubiquitous in molecular clouds and, as a result, a less tight correlation of
the CO J=1-0 line emission with H$\alpha$ luminosity is observed.

\section{Environmental effects on the dust and gas content}

One aspect related to the environment which we are going to address
is whether the global dust-to-gas ratio can be used as a metallicity indicator for cluster galaxies.  
We devote the rest of this section to investigate if the environment can enhance or inhibit the persistence
of dust and molecules in galaxies \citep{1986ApJ...310..660S,2008A&A...490..571F}.
The environmental effects are examined primarily in terms of HI deficiency which we have computed
for all galaxies in our sample following the prescription of \citet{1984AJ.....89..758H}. 

\begin{figure}
\includegraphics[width=\columnwidth]{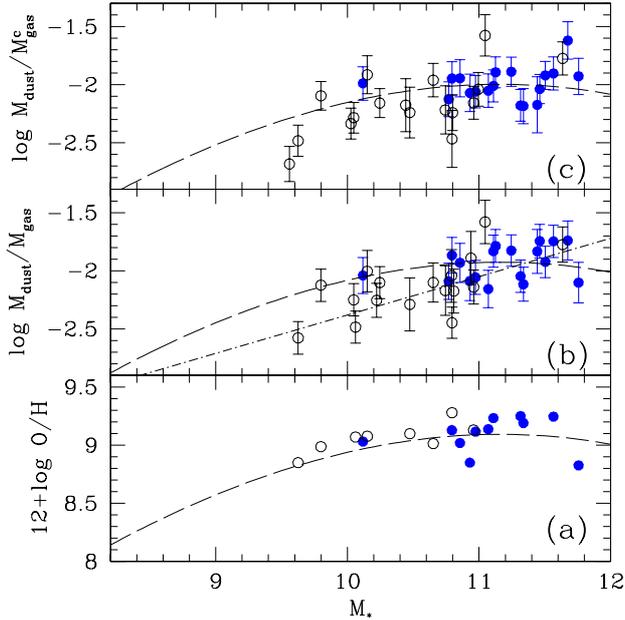}
\caption{In $(a)$ the mass-metallicity relation derived by \citet{2004ApJ...613..898T} 
(dashed line) is plotted together with data for galaxies in our sample with a known SDSS metallicity,  
given in Table~1.  In  panels $(b)$ and $(c)$
the  mass-metallicity relation has been scaled so that its peak reproduces  the average
dust-to-gas ratio for the M-sample. In panel $(b)$ M$_{H_2}^v$ is used to compute M$_{gas}$ and
the dot-dashed line is the best fit linear relation
to galaxies in the M- and S-sample.  In panel $(c)$ M$_{H_2}^c$ is used in computing M$_{tot}$. 
}
\label{fig:mm}
\end{figure}

In Figure~\ref{fig:mm}$(a)$ we show the consistency of
the metallicities of the central regions of galaxies in our sample with the mass-metallicity relation.
In  Figure~\ref{fig:mm}$(b)$ the dust-to-gas ratio is plotted as a function of stellar mass. 
The dashed line  shows the mass-metallicity relation scaled so that its peak reproduces
the average dust-to-gas ratio found for the M-sample, i.e. -1.9$\pm 0.15$ in log. This implies 
a dust-to-gas ratio of 0.013, a factor
1.8 higher than the value of 0.007 found for the Milky Way \citep{2007ApJ...663..866D}, 
whose central regions have metal abundances comparable to those of the M-sample, but only
slightly higher that the theoretical value of 0.01 \citep{2007ApJ...663..866D}.   
In panel $(c)$ we show the better agreement with the mass-metallicity relation if M$_{H_2}^c$
is used instead (i.e. a constant CO-to-H$_2$ conversion factor). In this case the discrepancy
between the average dust-to-gas ratio in the Milky Way and that of the M-sample is reduced. 

Galaxies in the S-sample lie mostly below the mass-metallicity relation both  in panel $(b)$
and $(c)$. For galaxies in the S-sample the dust-to-gas ratio is lower than what a linear scaling 
with the metallicity would predict, similar to what \citet{2009ApJ...701.1965M} find.
These results holds even if we consider HI deficient and non-HI deficient galaxies 
separately. Likely, this is a consequence that metal abundances and dust-to-gas ratios
sample different galaxy regions. The SDSS survey measures metal abundances at the galaxy center
while the dust-to-gas ratio in this paper is a quantity integrated over the whole disk.
Unenriched extended gaseous disks are more pronounced in low  mass blue galaxies. 
The average dust-to-gas ratio for the M- and S-sample together is   -2.1$\pm 0.3$ in log. 
The best linear fit to the M- and S-sample data for M$_d$/M$_{gas}$ as a function of M$_*$
reads:

\begin{equation}
{\hbox{log}} M_{dust}/M_{gas} = 0.26\times {\hbox{log}} M_* - 4.9  
\end{equation}

\noindent
This relation (with correlation coefficient $r=0.65$) is shown by the dot-dashed line in Figure~\ref{fig:mm}$(b)$.

\begin{figure*}
\centering
\includegraphics[width=14cm]{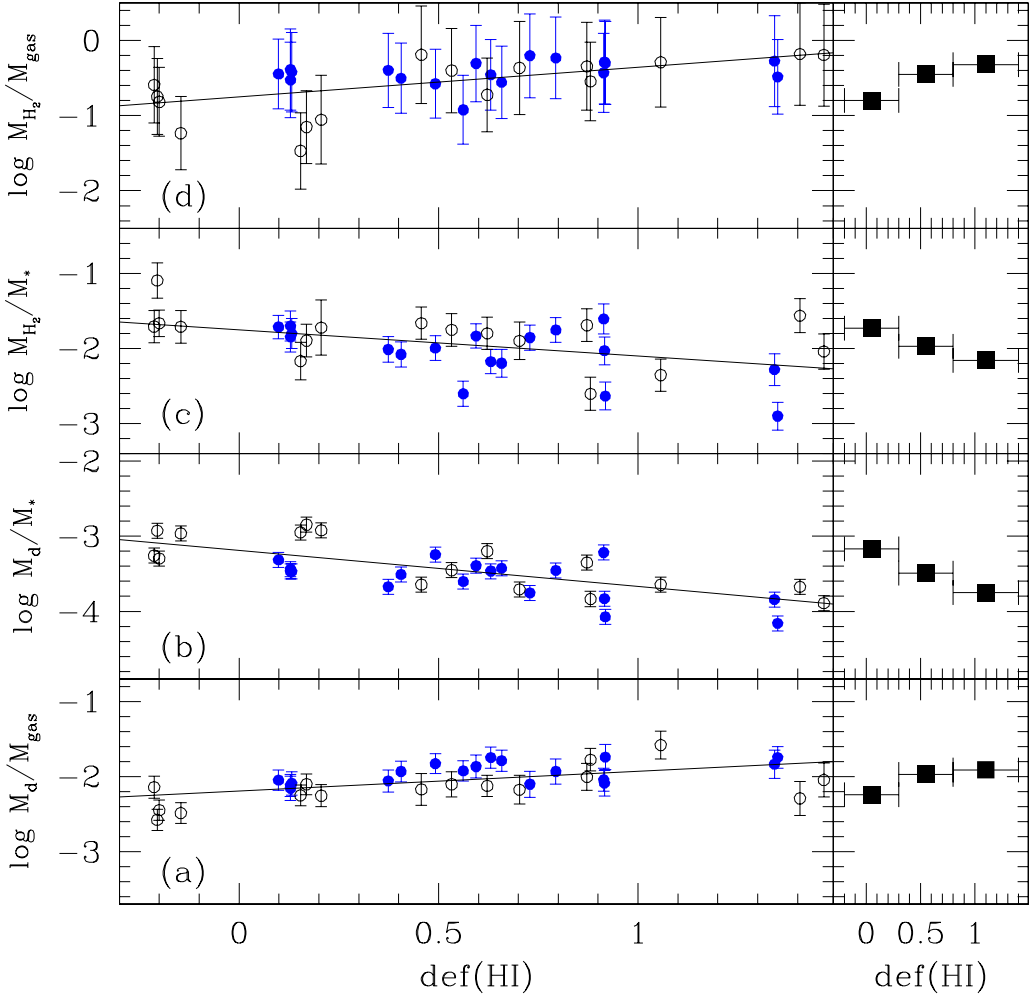}
\caption{The dust-to-gas ratio, dust-to-stellar mass ratio,   
molecular-to-stellar mass ratio and the molecular fraction
as a function of the HI deficiency. Panels to the left
show the data with symbols as in previous figures: filled circles are galaxies in
the M-sample,  open circles are galaxies in the S-sample. Continuous lines in each panel 
are best fits to linear relation between def(HI) and the variable shown on the y-axis. 
Panels to the right of the figure show mean values in three HI deficiency bins.}
\label{fig:def}
\end{figure*}

To check whether the HI deficiency affects the dust-to-gas ratio
we show this last variable as a function of the HI deficiency, def(HI), in the bottom panel of  
Figure~\ref{fig:def}. We can see that as the HI deficiency increases,  
the dust-to-gas ratio increases but the correlation is weak
(best fit slope is 0.26$\pm0.06$, $r=0.58$). 
This trend can be due to two possible effects: 
the first one is gas removal which affects the dust content less  because dust lies more
deeply in the potential well of the galaxy. The second one is  an evolutionary
effect: if the galaxy entering the cluster turns off gas infall, it uses its
gas reservoir to make more metals and stars. If this were the dominant effect,
galaxies with higher metallicity (with respect to the metallicities
predicted by the mass-metallicity relation) should be more HI deficient.
We do not find any correlation between the HI deficiency and the scatter in metallicity 
with respect to the mass-metallicity relation. Hence,
gas removal is responsible for the observed trend.
The small panels on the right side of Figure~\ref{fig:def} show the average values in 3 def(HI)
bins. It is important to note that the dust-to-gas mass ratio does not increase as the
galaxy becomes highly HI deficient, while between non-HI deficient galaxies and galaxies which
have a mild deficiency this ratio increases by a factor two. This finding suggests that
when the galaxy is highly disturbed the galaxy also becomes dust deficient. 
In terms of total mass, a small disturbance
mostly affects the outer HI envelope and does not much affect the total dust mass   
since  dust is  more segregated in the inner regions of the disk (both in height and radially). 
A similar conclusion has been reached by \citet{2012A&A...540A..52C} by comparing a sample
of field and cluster spirals. They find that the effect of dust removal from the star-forming disk of cluster 
galaxies is less strong than what is observed for the atomic gas. 

To investigate further whether some fraction of dust mass is lost as the galaxy moves through the cluster, 
we normalize the dust mass to the stellar mass. 
Dust removal is responsible for the correlation between the dust-to-stellar mass 
ratio and the HI deficiency shown in Figure~\ref{fig:def}$(b)$. Galaxies with a higher HI deficiency
have a lower dust-to-stellar mass ratio, only 25$\%$ of that observed in non HI-deficient galaxies.
The slope of the correlation is -0.48$\pm 0.08$ and the Pearson linear correlation coefficient is -0.72. 
Dust deficiency in HI deficient Virgo spirals has been pointed out already by \citet{2010A&A...518L..49C} in a spatially
resolved study. The authors show that the extent of the dust disk is significantly reduced in HI truncated  disk, due
to stripping by the cluster environment 
There is the possibility that dust removal could be more effective 
in low-metallicity, low-mass galaxies with shallow potential wells \citep{2006PASP..118..517B,2007MNRAS.380.1313B}
but the HI deficiency does not show any dependence on the stellar mass.

To analyze possible environmental effects on the molecular component we use the M$_{H_2}$-to-M$_*$ and
M$_{H_2}$-to-M$_{gas}$ ratios. 
We show these as a function of def(HI) in the top panels  of Figure~\ref{fig:def}.
The best fitting linear relation between log M$_{H_2}$/M$_*$ and def(HI) has a slope
of -0.35$\pm0.11$. 
The dependence on the HI deficiency of ratios involving the molecular mass, or the CO luminosity,  
is weaker (r$\sim 0.5$) than for the dust component.  
If confirmed by future data this correlation  implies that HI deficient galaxies are also H$_2$
deficient due to molecular gas removal or to quenching of molecule formation in  perturbed
gas. For our sample the M$_{H_2}$/M$_*$--def(HI) relation does not dependent on X$_{CO}$:
the slope of the correlation between the L$_{CO}$-to-M$_*$ ratio and the HI deficiency is also shallow 
(-0.28$\pm 0.09$) and consistent with that found for the M$_{H_2}$-to-M$_*$ ratio. 
The gas molecular fraction, M$_{H_2}$/M$_{gas}$, increases as the HI deficiency increases 
(Figure~\ref{fig:def}$(d)$) because the atomic gas mass is more heavily removed in the cluster
environment.  

As shown in Figure~\ref{fig:def}$(c)$, highly HI deficient galaxies
can quench or remove as much as 50-60$\%$ of their original molecular content. 
If we consider only the mapped galaxies in the M-sample the decrease of H$_2$ content is confirmed and becomes 
more extreme since the CO luminosity, as well as the H$_2$ mass, shows a steeper decrease with the HI 
deficiency parameter. A future comparison with a field galaxy 
sample is desirable to confirm this finding.  The cluster environment strips out part of the tenuous atomic ISM 
(e.g. outer disks and high latitude gas) more easily than dust 
and H$_2$ gas, which is more confined to the mid-plane and to the inner star forming disk. In terms of
total mass, dust and molecules are less affected by the environment than the HI gas.
Highly disturbed galaxies have considerably lower dust  content than their quiescent
counterparts  and higher molecular fractions.    

\begin{figure}
\includegraphics[width=\columnwidth]{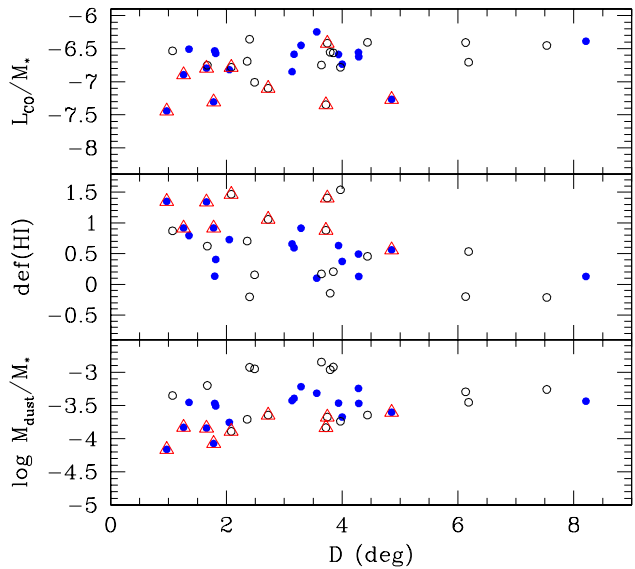}
\caption{The dust-stellar mass ratio, the HI deficiency and the ratio between the CO luminosity and the stellar
mass are shown  as a function of the
projected distance from the Virgo cluster center. Galaxies  are coded as in Figure~\ref{fig:f1}.
The open (red) triangles indicate galaxies of morphological type Sab or Sb.}
\label{fig:dist}
\end{figure}

Figure~\ref{fig:dist} shows possible dependencies of galaxy properties on the distance from the 
cluster center. Non-HI deficient galaxies avoid the cluster center, where a higher fraction of Sab and Sb (open
triangles)  can be found. The dispersion 
across the whole cluster is large because we are using projected distances.
The evidence that dust deficient galaxies and galaxies with a lower L$_{CO}$/M$_*$  
are  located in projection near the cluster center is marginal  (see bottom and top
panel of Figure~\ref{fig:dist}). 
  
\section{Summary}

   The relations between dust and gaseous masses have been examined in this paper for
   a magnitude limited sample of late type Virgo Cluster galaxies in the HeViCS fields.  
   The {\it Herschel Space Observatory} has detected cold dust emission for all galaxies in our sample 
   with the PACS and SPIRE detector at 100, 160, 250, 350 and 500~$\mu$m.  We summarize below the main results:
    
\begin{itemize}

\item
There is a tight correlation between the total CO J=1-0 flux and the cold dust emission.
Molecules are more closely related to cold dust rather than to dust heated by star formation or
to other galaxy integrated quantities such as blue luminosity, star formation  or stellar mass.  

\item
Dust mass correlates better with  the total (atomic+molecular) gas mass rather than with the atomic or
molecular gas alone. Infrared-dim galaxies are unable to convert much of their atomic
gas into molecular gas. Galaxies with a low molecular-to-total gas ratios have lower 
dust temperatures, are intrinsically faint and blue and a have a high gas-to-stellar mass ratios.  

\item
The dust-to-gas mass ratio typically decreases as the galaxy  stellar mass decreases, faster than predicted by
the mass-metallicity relation. We find a population of galaxies with a lower dust-to-gas ratio than what a linear 
scaling with metallicity would imply. Likely this is because the total gas
mass of a galaxy includes the unenriched extended disk. In fact these galaxies are HI rich with a less prominent 
stellar mass. 

\item
Galaxies with a mild HI-deficiency have a higher dust-to-gas ratios than non HI-deficient galaxies. 
This implies a higher efficiency of the
cluster stripping process on the HI gas rather than on the dust, which is more confined inside
the disk.  However, the cluster environment is able to remove some dust from the galaxy disk as well.
The dust mass per unit stellar mass decreases as the HI deficiency increases, and highly HI deficient 
galaxies can have up to  75$\%$ less dust than non HI deficient galaxies.

\item
There is evidence of molecular hydrogen quenching or stripping in HI deficient galaxies.
This evidence is found by considering the molecular gas per unit stellar mass. If its amplitude is 
confirmed by future data implies that  
molecules are missing in Virgo HI deficient spirals, but to a somewhat lesser extent than dust.
The molecular fraction increases as the HI deficiency increases because the cluster environment affects 
molecules to a lesser extent than atomic gas.

\item
Our results have been confirmed using various assumptions on the CO-to-H$_2$ conversion factor.
We cannot constrain possible dependencies of X$_{CO}$ from galaxy properties, such as the disk morphology or 
metallicity, due to the limited dynamical range of these parameters in the sample examined. 

\end{itemize}

\begin{acknowledgements}
The recent IRAM observations presented in this paper have benefited from research funding from the European 
Community's Seventh Framework Programme.
The research leading to these results has received funding from the Agenzia Spaziale Italiana 
(ASI-INAF agreements I/016/07/0 and I/009/10/0) and from the European Community's Seventh Framework Programme 
(/FP7/2007-2013/ under grant agreement No 229517).
C. V. received support from the ALMA-CONICYT Fund for the
Development of Chilean Astronomy (Project 31090013) and from the
Center of Excellence in Astrophysics and Associated Technologies (PBF06).
\end{acknowledgements}

\bibliography{astroph}

\begin{thebibliography}{69}
\expandafter\ifx\csname natexlab\endcsname\relax\def\natexlab#1{#1}\fi

\bibitem[{{Abdo} {et~al.}(2010){Abdo}, {Ackermann}, {Ajello}, {Baldini},
  {Ballet}, {Barbiellini}, {Bastieri}, {Baughman}, {Bechtol}, {Bellazzini},
  {Berenji}, {Bloom}, {Bonamente}, {Borgland}, {Bregeon}, {Brez}, {Brigida},
  {Bruel}, {Burnett}, {Buson}, {Caliandro}, {Cameron}, {Caraveo}, {Casandjian},
  {Cecchi}, {{\c C}elik}, {Chekhtman}, {Cheung}, {Chiang}, {Ciprini}, {Claus},
  {Cohen-Tanugi}, {Cominsky}, {Conrad}, {Dermer}, {de Palma}, {Digel}, {Silva},
  {Drell}, {Dubois}, {Dumora}, {Farnier}, {Favuzzi}, {Fegan}, {Focke},
  {Fortin}, {Frailis}, {Fukazawa}, {Funk}, {Fusco}, {Gargano}, {Gehrels},
  {Germani}, {Giavitto}, {Giebels}, {Giglietto}, {Giordano}, {Glanzman},
  {Godfrey}, {Grenier}, {Grondin}, {Grove}, {Guillemot}, {Guiriec}, {Harding},
  {Hayashida}, {Horan}, {Hughes}, {Jackson}, {J{\'o}hannesson}, {Johnson},
  {Johnson}, {Kamae}, {Katagiri}, {Kataoka}, {Kawai}, {Kerr}, {Kn{\"o}dlseder},
  {Kuss}, {Lande}, {Latronico}, {Lemoine-Goumard}, {Longo}, {Loparco}, {Lott},
  {Lovellette}, {Lubrano}, {Makeev}, {Mazziotta}, {McEnery}, {Meurer},
  {Michelson}, {Mitthumsiri}, {Mizuno}, {Monte}, {Monzani}, {Morselli},
  {Moskalenko}, {Murgia}, {Nolan}, {Norris}, {Nuss}, {Ohsugi}, {Okumura},
  {Omodei}, {Orlando}, {Ormes}, {Paneque}, {Pelassa}, {Pepe}, {Pesce-Rollins},
  {Piron}, {Porter}, {Rain{\`o}}, {Rando}, {Razzano}, {Reimer}, {Reimer},
  {Reposeur}, {Rodriguez}, {Ryde}, {Sadrozinski}, {Sanchez}, {Sander}, {Saz
  Parkinson}, {Sgr{\`o}}, {Siskind}, {Smith}, {Spandre}, {Spinelli}, {Starck},
  {Strickman}, {Strong}, {Suson}, {Takahashi}, {Tanaka}, {Thayer}, {Thayer},
  {Thompson}, {Tibaldo}, {Torres}, {Tosti}, {Tramacere}, {Uchiyama}, {Usher},
  {Vasileiou}, {Vilchez}, {Vitale}, {Waite}, {Wang}, {Winer}, {Wood}, {Ylinen},
  {Ziegler}, \& {Fermi/LAT Collaboration}}]{2010ApJ...710..133A}
{Abdo}, A.~A., {Ackermann}, M., {Ajello}, M., {et~al.} 2010, \apj, 710, 133

\bibitem[{{Auld} {et~al.}(2012){Auld} et al.}]{2012mnras submitted}
{Auld}, R., et al. 2012, \mnras, submitted

\bibitem[{{Bekki} {et~al.}(2002){Bekki}, {Couch}, \&
  {Shioya}}]{2002ApJ...577..651B}
{Bekki}, K., {Couch}, W.~J., \& {Shioya}, Y. 2002, \apj, 577, 651

\bibitem[{{Bendo} {et~al.}(2007){Bendo}, {Calzetti}, {Engelbracht},
  {Kennicutt}, {Meyer}, {Thornley}, {Walter}, {Dale}, {Li}, \&
  {Murphy}}]{2007MNRAS.380.1313B}
{Bendo}, G.~J., {Calzetti}, D., {Engelbracht}, C.~W., {et~al.} 2007, \mnras,
  380, 1313

\bibitem[{{Bendo} {et~al.}(2010{\natexlab{a}}){Bendo}, {Wilson}, {Pohlen},
  {Sauvage}, {Auld}, {Baes}, {Barlow}, {Bock}, {Boselli}, {Bradford}, {Buat},
  {Castro-Rodriguez}, {Chanial}, {Charlot}, {Ciesla}, {Clements}, {Cooray},
  {Cormier}, {Cortese}, {Davies}, {Dwek}, {Eales}, {Elbaz}, {Galametz},
  {Galliano}, {Gear}, {Glenn}, {Gomez}, {Griffin}, {Hony}, {Isaak}, {Levenson},
  {Lu}, {Madden}, {O'Halloran}, {Okumura}, {Oliver}, {Page}, {Panuzzo},
  {Papageorgiou}, {Parkin}, {Perez-Fournon}, {Rangwala}, {Rigby}, {Roussel},
  {Rykala}, {Sacchi}, {Schulz}, {Schirm}, {Smith}, {Spinoglio}, {Stevens},
  {Sundar}, {Symeonidis}, {Trichas}, {Vaccari}, {Vigroux}, {Wozniak}, {Wright},
  \& {Zeilinger}}]{2010A&A...518L..65B}
{Bendo}, G.~J., {Wilson}, C.~D., {Pohlen}, M., {et~al.} 2010{\natexlab{a}},
  \aap, 518, L65+

\bibitem[{{Bendo} {et~al.}(2010{\natexlab{b}}){Bendo}, {Wilson}, {Warren},
  {Brinks}, {Butner}, {Chanial}, {Clements}, {Courteau}, {Irwin}, {Israel},
  {Knapen}, {Leech}, {Matthews}, {M{\"u}hle}, {Petitpas}, {Serjeant}, {Tan},
  {Tilanus}, {Usero}, {Vaccari}, {van der Werf}, {Vlahakis}, {Wiegert}, \&
  {Zhu}}]{2010MNRAS.402.1409B}
{Bendo}, G.~J., {Wilson}, C.~D., {Warren}, B.~E., {et~al.} 2010{\natexlab{b}},
  \mnras, 402, 1409

\bibitem[{{Binggeli} {et~al.}(1985){Binggeli}, {Sandage}, \&
  {Tammann}}]{1985AJ.....90.1681B}
{Binggeli}, B., {Sandage}, A., \& {Tammann}, G.~A. 1985, \aj, 90, 1681

\bibitem[{{Bolatto} {et~al.}(2008){Bolatto}, {Leroy}, {Rosolowsky}, {Walter},
  \& {Blitz}}]{2008ApJ...686..948B}
{Bolatto}, A.~D., {Leroy}, A.~K., {Rosolowsky}, E., {Walter}, F., \& {Blitz},
  L. 2008, \apj, 686, 948

\bibitem[{{Boselli} {et~al.}(1995){Boselli}, {Casoli}, \&
  {Lequeux}}]{1995A&AS..110..521B}
{Boselli}, A., {Casoli}, F., \& {Lequeux}, J. 1995, \aaps, 110, 521

\bibitem[{{Boselli} \& {Gavazzi}(2006)}]{2006PASP..118..517B}
{Boselli}, A. \& {Gavazzi}, G. 2006, \pasp, 118, 517

\bibitem[{{Boselli} {et~al.}(2002){Boselli}, {Lequeux}, \&
  {Gavazzi}}]{2002A&A...384...33B}
{Boselli}, A., {Lequeux}, J., \& {Gavazzi}, G. 2002, \aap, 384, 33

\bibitem[{{Charlot} \& {Longhetti}(2001)}]{2001MNRAS.323..887C}
{Charlot}, S. \& {Longhetti}, M. 2001, \mnras, 323, 887

\bibitem[{{Chung} {et~al.}(2009{\natexlab{a}}){Chung}, {van Gorkom}, {Kenney},
  {Crowl}, \& {Vollmer}}]{2009AJ....138.1741C}
{Chung}, A., {van Gorkom}, J.~H., {Kenney}, J.~D.~P., {Crowl}, H., \&
  {Vollmer}, B. 2009{\natexlab{a}}, \aj, 138, 1741

\bibitem[{{Chung} {et~al.}(2009{\natexlab{b}}){Chung}, {Rhee}, {Kim}, {Yun},
  {Heyer}, \& {Young}}]{2009ApJS..184..199C}
{Chung}, E.~J., {Rhee}, M., {Kim}, H., {et~al.} 2009{\natexlab{b}}, \apjs, 184,
  199

\bibitem[{{Cortese} {et~al.}(2012){Cortese}, {Ciesla}, {Boselli}, {Bianchi},
  {Gomez}, {Smith}, {Bendo}, {Eales}, {Pohlen}, {Baes}, {Corbelli}, {Davies},
  {Hughes}, {Hunt}, {Madden}, {Pierini}, {di Serego Alighieri}, {Zibetti},
  {Boquien}, {Clements}, {Cooray}, {Galametz}, {Magrini}, {Pappalardo},
  {Spinoglio}, \& {Vlahakis}}]{2012A&A...540A..52C}
{Cortese}, L., {Ciesla}, L., {Boselli}, A., {et~al.} 2012, \aap, 540, A52

\bibitem[{{Cortese} {et~al.}(2010){Cortese}, {Davies}, {Pohlen}, {Baes},
  {Bendo}, {Bianchi}, {Boselli}, {de Looze}, {Fritz}, {Verstappen}, {Bomans},
  {Clemens}, {Corbelli}, {Dariush}, {di Serego Alighieri}, {Fadda},
  {Garcia-Appadoo}, {Gavazzi}, {Giovanardi}, {Grossi}, {Hughes}, {Hunt},
  {Jones}, {Madden}, {Pierini}, {Sabatini}, {Smith}, {Vlahakis}, {Xilouris}, \&
  {Zibetti}}]{2010A&A...518L..49C}
{Cortese}, L., {Davies}, J.~I., {Pohlen}, M., {et~al.} 2010, \aap, 518, L49+

\bibitem[{{Dame} {et~al.}(2001){Dame}, {Hartmann}, \&
  {Thaddeus}}]{2001ApJ...547..792D}
{Dame}, T.~M., {Hartmann}, D., \& {Thaddeus}, P. 2001, \apj, 547, 792

\bibitem[{{Davies} {et~al.}(2010){Davies}, {Baes}, {Bendo}, {Bianchi},
  {Bomans}, {Boselli}, {Clemens}, {Corbelli}, {Cortese}, {Dariush}, {de Looze},
  {di Serego Alighieri}, {Fadda}, {Fritz}, {Garcia-Appadoo}, {Gavazzi},
  {Giovanardi}, {Grossi}, {Hughes}, {Hunt}, {Jones}, {Madden}, {Pierini},
  {Pohlen}, {Sabatini}, {Smith}, {Verstappen}, {Vlahakis}, {Xilouris}, \&
  {Zibetti}}]{2010A&A...518L..48D}
{Davies}, J.~I., {Baes}, M., {Bendo}, G.~J., {et~al.} 2010, \aap, 518, L48+

\bibitem[{{Davies} {et~al.}(2012){Davies}, {Bianchi}, {Cortese}, {Auld},
  {Baes}, {Bendo}, {Boselli}, {Ciesla}, {Clemens}, {Corbelli}, {de Looze},
  {Alighieri}, {Fritz}, {Gavazzi}, {Pappalardo}, {Grossi}, {Hunt}, {Madden},
  {Magrini}, {Pohlen}, {Smith}, {Verstappen}, \&
  {Vlahakis}}]{2012MNRAS.419.3505D}
{Davies}, J.~I., {Bianchi}, S., {Cortese}, L., {et~al.} 2012, \mnras, 419, 3505

\bibitem[{{de Vaucouleurs} {et~al.}(1995){de Vaucouleurs}, {de Vaucouleurs},
  {Corwin}, {Buta}, {Paturel}, \& {Fouque}}]{1995yCat.7155....0D}
{de Vaucouleurs}, G., {de Vaucouleurs}, A., {Corwin}, H.~G., {et~al.} 1995,
  VizieR Online Data Catalog, 7155, 0

\bibitem[{{Di Matteo} {et~al.}(2007){Di Matteo}, {Combes}, {Melchior}, \&
  {Semelin}}]{2007A&A...468...61D}
{Di Matteo}, P., {Combes}, F., {Melchior}, A., \& {Semelin}, B. 2007, \aap,
  468, 61

\bibitem[{{Draine}(2003)}]{2003ARA&A..41..241D}
{Draine}, B.~T. 2003, \araa, 41, 241

\bibitem[{{Draine} \& {Bertoldi}(1996)}]{1996ApJ...468..269D}
{Draine}, B.~T. \& {Bertoldi}, F. 1996, \apj, 468, 269

\bibitem[{{Draine} {et~al.}(2007){Draine}, {Dale}, {Bendo}, {Gordon}, {Smith},
  {Armus}, {Engelbracht}, {Helou}, {Kennicutt}, {Li}, {Roussel}, {Walter},
  {Calzetti}, {Moustakas}, {Murphy}, {Rieke}, {Bot}, {Hollenbach}, {Sheth}, \&
  {Teplitz}}]{2007ApJ...663..866D}
{Draine}, B.~T., {Dale}, D.~A., {Bendo}, G., {et~al.} 2007, \apj, 663, 866

\bibitem[{{Elmegreen}(1993)}]{1993ApJ...411..170E}
{Elmegreen}, B.~G. 1993, \apj, 411, 170

\bibitem[{{Fumagalli} \& {Gavazzi}(2008)}]{2008A&A...490..571F}
{Fumagalli}, M. \& {Gavazzi}, G. 2008, \aap, 490, 571

\bibitem[{{Fumagalli} {et~al.}(2009){Fumagalli}, {Krumholz}, {Prochaska},
  {Gavazzi}, \& {Boselli}}]{2009ApJ...697.1811F}
{Fumagalli}, M., {Krumholz}, M.~R., {Prochaska}, J.~X., {Gavazzi}, G., \&
  {Boselli}, A. 2009, \apj, 697, 1811

\bibitem[{{Galametz} {et~al.}(2011){Galametz}, {Madden}, {Galliano}, {Hony},
  {Bendo}, \& {Sauvage}}]{2011A&A...532A..56G}
{Galametz}, M., {Madden}, S.~C., {Galliano}, F., {et~al.} 2011, \aap, 532, A56

\bibitem[{{Gavazzi} \& {Boselli}(1996)}]{1996ApL&C..35....1G}
{Gavazzi}, G. \& {Boselli}, A. 1996, Astrophysical Letters Communications, 35,
  1

\bibitem[{{Gavazzi} {et~al.}(2003){Gavazzi}, {Boselli}, {Donati}, {Franzetti},
  \& {Scodeggio}}]{2003A&A...400..451G}
{Gavazzi}, G., {Boselli}, A., {Donati}, A., {Franzetti}, P., \& {Scodeggio}, M.
  2003, \aap, 400, 451

\bibitem[{{Gavazzi} {et~al.}(1999){Gavazzi}, {Boselli}, {Scodeggio}, {Pierini},
  \& {Belsole}}]{1999MNRAS.304..595G}
{Gavazzi}, G., {Boselli}, A., {Scodeggio}, M., {Pierini}, D., \& {Belsole}, E.
  1999, \mnras, 304, 595

\bibitem[{{Giovanelli} \& {Haynes}(1985)}]{1985ApJ...292..404G}
{Giovanelli}, R. \& {Haynes}, M.~P. 1985, \apj, 292, 404

\bibitem[{{Glover} \& {Mac Low}(2011)}]{2011MNRAS.412..337G}
{Glover}, S.~C.~O. \& {Mac Low}, M. 2011, \mnras, 412, 337

\bibitem[{{Gould} \& {Salpeter}(1963)}]{1963ApJ...138..393G}
{Gould}, R.~J. \& {Salpeter}, E.~E. 1963, \apj, 138, 393

\bibitem[{{Gratier} {et~al.}(2010){Gratier}, {Braine}, {Rodriguez-Fernandez},
  {Israel}, {Schuster}, {Brouillet}, \& {Gardan}}]{2010A&A...512A..68G}
{Gratier}, P., {Braine}, J., {Rodriguez-Fernandez}, N.~J., {et~al.} 2010, \aap,
  512, A68+

\bibitem[{{Griffin} {et~al.}(2010){Griffin}, {Abergel}, {Abreu}, {Ade},
  {Andr{\'e}}, {Augueres}, {Babbedge}, {Bae}, {Baillie}, {Baluteau}, {Barlow},
  {Bendo}, {Benielli}, {Bock}, {Bonhomme}, {Brisbin}, {Brockley-Blatt},
  {Caldwell}, {Cara}, {Castro-Rodriguez}, {Cerulli}, {Chanial}, {Chen},
  {Clark}, {Clements}, {Clerc}, {Coker}, {Communal}, {Conversi}, {Cox},
  {Crumb}, {Cunningham}, {Daly}, {Davis}, {de Antoni}, {Delderfield}, {Devin},
  {di Giorgio}, {Didschuns}, {Dohlen}, {Donati}, {Dowell}, {Dowell}, {Duband},
  {Dumaye}, {Emery}, {Ferlet}, {Ferrand}, {Fontignie}, {Fox}, {Franceschini},
  {Frerking}, {Fulton}, {Garcia}, {Gastaud}, {Gear}, {Glenn}, {Goizel},
  {Griffin}, {Grundy}, {Guest}, {Guillemet}, {Hargrave}, {Harwit}, {Hastings},
  {Hatziminaoglou}, {Herman}, {Hinde}, {Hristov}, {Huang}, {Imhof}, {Isaak},
  {Israelsson}, {Ivison}, {Jennings}, {Kiernan}, {King}, {Lange}, {Latter},
  {Laurent}, {Laurent}, {Leeks}, {Lellouch}, {Levenson}, {Li}, {Li},
  {Lilienthal}, {Lim}, {Liu}, {Lu}, {Madden}, {Mainetti}, {Marliani}, {McKay},
  {Mercier}, {Molinari}, {Morris}, {Moseley}, {Mulder}, {Mur}, {Naylor},
  {Nguyen}, {O'Halloran}, {Oliver}, {Olofsson}, {Olofsson}, {Orfei}, {Page},
  {Pain}, {Panuzzo}, {Papageorgiou}, {Parks}, {Parr-Burman}, {Pearce},
  {Pearson}, {P{\'e}rez-Fournon}, {Pinsard}, {Pisano}, {Podosek}, {Pohlen},
  {Polehampton}, {Pouliquen}, {Rigopoulou}, {Rizzo}, {Roseboom}, {Roussel},
  {Rowan-Robinson}, {Rownd}, {Saraceno}, {Sauvage}, {Savage}, {Savini},
  {Sawyer}, {Scharmberg}, {Schmitt}, {Schneider}, {Schulz}, {Schwartz},
  {Shafer}, {Shupe}, {Sibthorpe}, {Sidher}, {Smith}, {Smith}, {Smith},
  {Spencer}, {Stobie}, {Sudiwala}, {Sukhatme}, {Surace}, {Stevens}, {Swinyard},
  {Trichas}, {Tourette}, {Triou}, {Tseng}, {Tucker}, {Turner}, {Vaccari},
  {Valtchanov}, {Vigroux}, {Virique}, {Voellmer}, {Walker}, {Ward}, {Waskett},
  {Weilert}, {Wesson}, {White}, {Whitehouse}, {Wilson}, {Winter}, {Woodcraft},
  {Wright}, {Xu}, {Zavagno}, {Zemcov}, {Zhang}, \&
  {Zonca}}]{2010A&A...518L...3G}
{Griffin}, M.~J., {Abergel}, A., {Abreu}, A., {et~al.} 2010, \aap, 518, L3+

\bibitem[{{Haynes} \& {Giovanelli}(1984)}]{1984AJ.....89..758H}
{Haynes}, M.~P. \& {Giovanelli}, R. 1984, \aj, 89, 758

\bibitem[{{Hirashita} {et~al.}(2002){Hirashita}, {Tajiri}, \&
  {Kamaya}}]{2002A&A...388..439H}
{Hirashita}, H., {Tajiri}, Y.~Y., \& {Kamaya}, H. 2002, \aap, 388, 439

\bibitem[{{Hollenbach} {et~al.}(1971){Hollenbach}, {Werner}, \&
  {Salpeter}}]{1971ApJ...163..165H}
{Hollenbach}, D.~J., {Werner}, M.~W., \& {Salpeter}, E.~E. 1971, \apj, 163, 165
  
\bibitem[{{Israel}(1997{\natexlab{a}})}]{1997A&A...317...65I}
{Israel}, F.~P. 1997{\natexlab{a}}, \aap, 317, 65

\bibitem[{{Israel}(1997{\natexlab{b}})}]{1997A&A...328..471I}
{Israel}, F.~P. 1997{\natexlab{b}}, \aap, 328, 471

\bibitem[{{Kenney} \& {Young}(1989)}]{1989ApJ...344..171K}
{Kenney}, J.~D.~P. \& {Young}, J.~S. 1989, \apj, 344, 171

\bibitem[{{Kewley} \& {Ellison}(2008)}]{2008ApJ...681.1183K}
{Kewley}, L.~J. \& {Ellison}, S.~L. 2008, \apj, 681, 1183

\bibitem[{{Knapp} {et~al.}(1987){Knapp}, {Helou}, \&
  {Stark}}]{1987AJ.....94...54K}
{Knapp}, G.~R., {Helou}, G., \& {Stark}, A.~A. 1987, \aj, 94, 54

\bibitem[{{Kuno} {et~al.}(2007){Kuno}, {Sato}, {Nakanishi}, {Hirota}, {Tosaki},
  {Shioya}, {Sorai}, {Nakai}, {Nishiyama}, \&
  {Vila-Vilar{\'o}}}]{2007PASJ...59..117K}
{Kuno}, N., {Sato}, N., {Nakanishi}, H., {et~al.} 2007, \pasj, 59, 117

\bibitem[{{Larson} {et~al.}(1980){Larson}, {Tinsley}, \&
  {Caldwell}}]{1980ApJ...237..692L}
{Larson}, R.~B., {Tinsley}, B.~M., \& {Caldwell}, C.~N. 1980, \apj, 237, 692

\bibitem[{{Leroy} {et~al.}(2007){Leroy}, {Bolatto}, {Stanimirovic}, {Mizuno},
  {Israel}, \& {Bot}}]{2007ApJ...658.1027L}
{Leroy}, A., {Bolatto}, A., {Stanimirovic}, S., {et~al.} 2007, \apj, 658, 1027

\bibitem[{{Leroy} {et~al.}(2011){Leroy}, {Bolatto}, {Gordon}, {Sandstrom},
  {Gratier}, {Rosolowsky}, {Engelbracht}, {Mizuno}, {Corbelli}, {Fukui}, \&
  {Kawamura}}]{2011ApJ...737...12L}
{Leroy}, A.~K., {Bolatto}, A., {Gordon}, K., {et~al.} 2011, \apj, 737, 12

\bibitem[{{Leroy} {et~al.}(2009){Leroy}, {Walter}, {Bigiel}, {Usero}, {Weiss},
  {Brinks}, {de Blok}, {Kennicutt}, {Schuster}, {Kramer}, {Wiesemeyer}, \&
  {Roussel}}]{2009AJ....137.4670L}
{Leroy}, A.~K., {Walter}, F., {Bigiel}, F., {et~al.} 2009, \aj, 137, 4670

\bibitem[{{Lisenfeld} \& {Ferrara}(1998)}]{1998ApJ...496..145L}
{Lisenfeld}, U. \& {Ferrara}, A. 1998, \apj, 496, 145

\bibitem[{{Magrini} {et~al.}(2011){Magrini}, {Bianchi}, {Corbelli}, {Cortese},
  {Hunt}, {Smith}, {Vlahakis}, {Davies}, {Bendo}, {Baes}, {Boselli}, {Clemens},
  {Casasola}, {de Looze}, {Fritz}, {Giovanardi}, {Grossi}, {Hughes}, {Madden},
  {Pappalardo}, {Pohlen}, {di Serego Alighieri}, \&
  {Verstappen}}]{2011A&A...535A..13M}
{Magrini}, L., {Bianchi}, S., {Corbelli}, E., {et~al.} 2011, \aap, 535, A13

\bibitem[{{Mannucci} {et~al.}(2005){Mannucci}, {Della Valle}, {Panagia},
  {Cappellaro}, {Cresci}, {Maiolino}, {Petrosian}, \&
  {Turatto}}]{2005A&A...433..807M}
{Mannucci}, F., {Della Valle}, M., {Panagia}, N., {et~al.} 2005, \aap, 433, 807

\bibitem[{{Mu{\~n}oz-Mateos} {et~al.}(2009){Mu{\~n}oz-Mateos}, {Gil de Paz},
  {Boissier}, {Zamorano}, {Dale}, {P{\'e}rez-Gonz{\'a}lez}, {Gallego},
  {Madore}, {Bendo}, {Thornley}, {Draine}, {Boselli}, {Buat}, {Calzetti},
  {Moustakas}, \& {Kennicutt}}]{2009ApJ...701.1965M}
{Mu{\~n}oz-Mateos}, J.~C., {Gil de Paz}, A., {Boissier}, S., {et~al.} 2009,
  \apj, 701, 1965

\bibitem[{{Murray}(2011)}]{2011ApJ...729..133M}
{Murray}, N. 2011, \apj, 729, 133

\bibitem[{{Nilson}(1973)}]{1973ugcg.book.....N}
{Nilson}, P. 1973, {Uppsala general catalogue of galaxies}, ed. {Nilson, P.}

\bibitem[{{Pappalardo} {et~al.}(2012){Pappalardo} et al.}]{2012A&A to be submitted}
{Pappalardo}, C., et al. 2012, \aap, to be submitted

\bibitem[{{Pilbratt} {et~al.}(2010){Pilbratt}, {Riedinger}, {Passvogel},
  {Crone}, {Doyle}, {Gageur}, {Heras}, {Jewell}, {Metcalfe}, {Ott}, \&
  {Schmidt}}]{2010A&A...518L...1P}
{Pilbratt}, G.~L., {Riedinger}, J.~R., {Passvogel}, T., {et~al.} 2010, \aap,
  518, L1+

\bibitem[{{Planck Collaboration} {et~al.}(2011){Planck Collaboration},
  {Abergel}, {Ade}, {Aghanim}, {Arnaud}, {Ashdown}, {Aumont}, {Baccigalupi},
  {Balbi}, {Banday}, {Barreiro}, {Bartlett}, {Battaner}, {Benabed},
  {Beno{\^i}t}, {Bernard}, {Bersanelli}, {Bhatia}, {Bock}, {Bonaldi}, {Bond},
  {Borrill}, {Bouchet}, {Boulanger}, {Bucher}, {Burigana}, {Cabella},
  {Cardoso}, {Catalano}, {Cay{\'o}n}, {Challinor}, {Chamballu}, {Chiang},
  {Chiang}, {Christensen}, {Clements}, {Colombi}, {Couchot}, {Coulais},
  {Crill}, {Cuttaia}, {Danese}, {Davies}, {Davis}, {de Bernardis}, {de
  Gasperis}, {de Rosa}, {de Zotti}, {Delabrouille}, {Delouis}, {D{\'e}sert},
  {Dickinson}, {Dobashi}, {Donzelli}, {Dor{\'e}}, {D{\"o}rl}, {Douspis},
  {Dupac}, {Efstathiou}, {En{\ss}lin}, {Eriksen}, {Finelli}, {Forni},
  {Frailis}, {Franceschi}, {Galeotta}, {Ganga}, {Giard}, {Giardino},
  {Giraud-H{\'e}raud}, {Gonz{\'a}lez-Nuevo}, {G{\'o}rski}, {Gratton},
  {Gregorio}, {Gruppuso}, {Guillet}, {Hansen}, {Harrison},
  {Henrot-Versill{\'e}}, {Herranz}, {Hildebrandt}, {Hivon}, {Hobson}, {Holmes},
  {Hovest}, {Hoyland}, {Huffenberger}, {Jaffe}, {Jones}, {Jones}, {Juvela},
  {Keih{\"a}nen}, {Keskitalo}, {Kisner}, {Kneissl}, {Knox}, {Kurki-Suonio},
  {Lagache}, {Lamarre}, {Lasenby}, {Laureijs}, {Lawrence}, {Leach}, {Leonardi},
  {Leroy}, {Linden-V{\o}rnle}, {L{\'o}pez-Caniego}, {Lubin},
  {Mac{\'{\i}}as-P{\'e}rez}, {MacTavish}, {Maffei}, {Mandolesi}, {Mann},
  {Maris}, {Marshall}, {Martin}, {Mart{\'{\i}}nez-Gonz{\'a}lez}, {Masi},
  {Matarrese}, {Matthai}, {Mazzotta}, {McGehee}, {Meinhold}, {Melchiorri},
  {Mendes}, {Mennella}, {Mitra}, {Miville-Desch{\^e}nes}, {Moneti}, {Montier},
  {Morgante}, {Mortlock}, {Munshi}, {Murphy}, {Naselsky}, {Natoli},
  {Netterfield}, {N{\o}rgaard-Nielsen}, {Noviello}, {Novikov}, {Novikov},
  {Osborne}, {Pajot}, {Paladini}, {Pasian}, {Patanchon}, {Perdereau},
  {Perotto}, {Perrotta}, {Piacentini}, {Piat}, {Plaszczynski}, {Pointecouteau},
  {Polenta}, {Ponthieu}, {Poutanen}, {Pr{\'e}zeau}, {Prunet}, {Puget}, {Reach},
  {Rebolo}, {Reinecke}, {Renault}, {Ricciardi}, {Riller}, {Ristorcelli},
  {Rocha}, {Rosset}, {Rubi{\~n}o-Mart{\'{\i}}n}, {Rusholme}, {Sandri},
  {Santos}, {Savini}, {Scott}, {Seiffert}, {Shellard}, {Smoot}, {Starck},
  {Stivoli}, {Stolyarov}, {Sudiwala}, {Sygnet}, {Tauber}, {Terenzi},
  {Toffolatti}, {Tomasi}, {Torre}, {Tristram}, {Tuovinen}, {Umana},
  {Valenziano}, {Verstraete}, {Vielva}, {Villa}, {Vittorio}, {Wade}, {Wandelt},
  {Yvon}, {Zacchei}, \& {Zonca}}]{2011A&A...536A..25P}
{Planck Collaboration}, {Abergel}, A., {Ade}, P.~A.~R., {et~al.} 2011, \aap,
  536, A25

\bibitem[{{Poglitsch} {et~al.}(2010){Poglitsch}, {Waelkens}, {Geis},
  {Feuchtgruber}, {Vandenbussche}, {Rodriguez}, {Krause}, {Renotte}, {van
  Hoof}, {Saraceno}, {Cepa}, {Kerschbaum}, {Agn{\`e}se}, {Ali}, {Altieri},
  {Andreani}, {Augueres}, {Balog}, {Barl}, {Bauer}, {Belbachir}, {Benedettini},
  {Billot}, {Boulade}, {Bischof}, {Blommaert}, {Callut}, {Cara}, {Cerulli},
  {Cesarsky}, {Contursi}, {Creten}, {De Meester}, {Doublier}, {Doumayrou},
  {Duband}, {Exter}, {Genzel}, {Gillis}, {Gr{\"o}zinger}, {Henning},
  {Herreros}, {Huygen}, {Inguscio}, {Jakob}, {Jamar}, {Jean}, {de Jong},
  {Katterloher}, {Kiss}, {Klaas}, {Lemke}, {Lutz}, {Madden}, {Marquet},
  {Martignac}, {Mazy}, {Merken}, {Montfort}, {Morbidelli}, {M{\"u}ller},
  {Nielbock}, {Okumura}, {Orfei}, {Ottensamer}, {Pezzuto}, {Popesso},
  {Putzeys}, {Regibo}, {Reveret}, {Royer}, {Sauvage}, {Schreiber}, {Stegmaier},
  {Schmitt}, {Schubert}, {Sturm}, {Thiel}, {Tofani}, {Vavrek}, {Wetzstein},
  {Wieprecht}, \& {Wiezorrek}}]{2010A&A...518L...2P}
{Poglitsch}, A., {Waelkens}, C., {Geis}, N., {et~al.} 2010, \aap, 518, L2+

\bibitem[{{Rudolph} {et~al.}(2006){Rudolph}, {Fich}, {Bell}, {Norsen},
  {Simpson}, {Haas}, \& {Erickson}}]{2006ApJS..162..346R}
{Rudolph}, A.~L., {Fich}, M., {Bell}, G.~R., {et~al.} 2006, \apjs, 162, 346

\bibitem[{{Shetty} {et~al.}(2011){Shetty}, {Glover}, {Dullemond}, {Ostriker},
  {Harris}, \& {Klessen}}]{2011MNRAS.415.3253S}
{Shetty}, R., {Glover}, S.~C., {Dullemond}, C.~P., {et~al.} 2011, \mnras, 415,
  3253

\bibitem[{{Smith} {et~al.}(2010){Smith}, {Vlahakis}, {Baes}, {Bendo},
  {Bianchi}, {Bomans}, {Boselli}, {Clemens}, {Corbelli}, {Cortese}, {Dariush},
  {Davies}, {de Looze}, {di Serego Alighieri}, {Fadda}, {Fritz},
  {Garcia-Appadoo}, {Gavazzi}, {Giovanardi}, {Grossi}, {Hughes}, {Hunt},
  {Jones}, {Madden}, {Pierini}, {Pohlen}, {Sabatini}, {Verstappen}, {Xilouris},
  \& {Zibetti}}]{2010A&A...518L..51S}
{Smith}, M.~W.~L., {Vlahakis}, C., {Baes}, M., {et~al.} 2010, \aap, 518, L51+

\bibitem[{{Solomon} \& {Sage}(1988)}]{1988ApJ...334..613S}
{Solomon}, P.~M. \& {Sage}, L.~J. 1988, \apj, 334, 613

\bibitem[{{Stark} {et~al.}(1986){Stark}, {Knapp}, {Bally}, {Wilson}, {Penzias},
  \& {Rowe}}]{1986ApJ...310..660S}
{Stark}, A.~A., {Knapp}, G.~R., {Bally}, J., {et~al.} 1986, \apj, 310, 660

\bibitem[{{Strong} \& {Mattox}(1996)}]{1996A&A...308L..21S}
{Strong}, A.~W. \& {Mattox}, J.~R. 1996, \aap, 308, L21

\bibitem[{{Tremonti} {et~al.}(2004){Tremonti}, {Heckman}, {Kauffmann},
  {Brinchmann}, {Charlot}, {White}, {Seibert}, {Peng}, {Schlegel}, {Uomoto},
  {Fukugita}, \& {Brinkmann}}]{2004ApJ...613..898T}
{Tremonti}, C.~A., {Heckman}, T.~M., {Kauffmann}, G., {et~al.} 2004, \apj, 613,
  898

\bibitem[{{Vollmer} {et~al.}(2008){Vollmer}, {Braine}, {Pappalardo}, \&
  {Hily-Blant}}]{2008A&A...491..455V}
{Vollmer}, B., {Braine}, J., {Pappalardo}, C., \& {Hily-Blant}, P. 2008, \aap,
  491, 455

\bibitem[{{Wilson}(1995)}]{1995ApJ...448L..97W}
{Wilson}, C.~D. 1995, \apjl, 448, L97+

\bibitem[{{Young}(1999)}]{1999ApJ...514L..87Y}
{Young}, J.~S. 1999, \apjl, 514, L87

\bibitem[{{Young} {et~al.}(1995){Young}, {Xie}, {Tacconi}, {Knezek}, {Viscuso},
  {Tacconi-Garman}, {Scoville}, {Schneider}, {Schloerb}, {Lord}, {Lesser},
  {Kenney}, {Huang}, {Devereux}, {Claussen}, {Case}, {Carpenter}, {Berry}, \&
  {Allen}}]{1995ApJS...98..219Y}
{Young}, J.~S., {Xie}, S., {Tacconi}, L., {et~al.} 1995, \apjs, 98, 219

\end{thebibliography}

\end{document}